\documentclass[%
 reprint,
groupedaddress,
 amsmath,amssymb,
 aps,
 prx,
]{revtex4-2}

\usepackage[T1]{fontenc}
\usepackage[utf8]{inputenc} 
\usepackage{appendix}
\usepackage{dcolumn}
\usepackage{bm}
\usepackage{csquotes}
\usepackage{booktabs}
\usepackage{graphicx} 
\usepackage{amsmath}
\usepackage{comment}
\usepackage{hyperref}
\usepackage{xcolor}
\usepackage{soul}

\usepackage[normalem]{ulem}

\newcommand{\Ec}{\mathcal{E}}
\newcommand{\Vc}{\mathcal{V}}

\newcommand{\de}{d_{e}}

\newcommand{\oij}{\omega_{ij}}

\begin{document}
\title{Reducibility of higher-order to pairwise interactions: Social impact models on hypergraphs}

\author{Jaume Llabr\'es}
\affiliation{Institute for Cross-disciplinary Physics and Complex Systems IFISC (CSIC-UIB), Campus Universitat Illes Balears, 07122 Palma de Mallorca, Spain.}
 
\author{Ra\'ul Toral}
\affiliation{Institute for Cross-disciplinary Physics and Complex Systems IFISC (CSIC-UIB), Campus Universitat Illes Balears, 07122 Palma de Mallorca, Spain.}
 
\author{Maxi San Miguel}
\affiliation{Institute for Cross-disciplinary Physics and Complex Systems IFISC (CSIC-UIB), Campus Universitat Illes Balears, 07122 Palma de Mallorca, Spain.}
 
\author{Federico Vazquez}
\affiliation{Institute for Cross-disciplinary Physics and Complex Systems IFISC (CSIC-UIB), Campus Universitat Illes Balears, 07122 Palma de Mallorca, Spain,}

\affiliation{Instituto de Cálculo, FCEN, Universidad de Buenos Aires and CONICET, C1428EGA Buenos Aires, Argentina.}

\date{\today}

\begin{abstract}
We show that a general class of node-update social impact models with higher-order interactions on hypergraphs can be exactly mapped to an equivalent model with pairwise interactions on a weighted projected network. This mapping preserves the microscopic probabilities of changing the state of the nodes. As a particular case, we introduce hypergraph-voter models, for which we compute the weights of the projected network, both analytically and numerically, across several hypergraph ensembles, and we characterize their ordering dynamics through simulations of both higher-order and reduced pairwise dynamics. For a linear social impact function (\emph{hypergraph-linear voter model}) the weights of the projected network are static (state-independent), allowing us to develop a pair approximation that describes with accuracy the time evolution of macroscopic observables, which turn out to be independent of those weights. The macroscopic dynamics is thus equivalent to that of the standard voter model on the unweighted projected network. For a power-law social impact function (\emph{hypergraph-nonlinear voter model}) the weights of the projected network depend on the instantaneous system configuration. Nevertheless, the nonlinear voter model on the unweighted projected network still reproduces the main macroscopic trends for well connected hypergraphs.
\end{abstract} 

\maketitle

\section{Introduction}
Over the last decade, it has been emphasized that many biological, social, technological and physical systems cannot be fully described by pairwise (PW) dyadic interactions alone. In many cases, the fundamental units of interaction are often \emph{polyadic}, involving three or more entities. Therefore, relying solely on pairwise interactions may thus obscure important features of social dynamics~\cite{LambiotteNature,Moreno2020, Battiston2020,Modelling_Lambiotte,Battiston2025}. Hypergraphs and simplicial complexes provide natural mathematical frameworks to encode such higher-order (HO) interactions and to explore the rich collective phenomena that they generate~\cite{Bianconi_2021}. Unlike standard complex networks—where nodes represent agents and edges represent dyadic interactions—hypergraphs allow hyperedges containing multiple nodes, each hyperedge corresponding to an interacting group. Recent reviews have shown that incorporating hyperedges in the description of a system profoundly alters both its structural and dynamical properties, while highlighting key open theoretical challenges~\cite{Battiston2020, Battiston_Nature, Arruda2024}. Simplicial-complex models have also been used to describe tightly bound groups, high-dimensional social structures \cite{Kee01012013}, and the ways in which group members acquire and accumulate information~\cite{Greening-2015}.

However, not all group or many-agent interactions require an explicit hypergraph representation. Certain social processes—such as multiple exposures inducing social conformity~\cite{Asch1951}—can be adequately captured within a standard complex networks framework. For instance, phenomena of \textit{complex contagion}~\cite{Centola2007,Centola2018}, 
including adoption of innovations or propagation of rumors, are often described by models inspired in the threshold model by Granovetter~\cite{Granovetter,Watts} in which, in a complex network representation, adoption requires a minimum number of neighboring adopters. It is known that complex contagion models often exhibit a cascade mechanism leading to a discontinuous first-order transition~\cite{Watts,Gleeson}, in contrast to the typically continuous transitions found in dyadic epidemic models~\cite{Romualdo}. Yet similar discontinuous transitions have also been attributed to HO interactions in hypergraphs and simplicial complexes~\cite{Iacopini2019,Moreno2020}. A second example is the nonlinear voter model on a complex network~\cite{Ramirez2024,Tobias2025,Llabres2025}, in which the update of the state on an agent depends on the joint configuration of all its neighbors, effectively capturing the simultaneous influence of group interactions. In fact, the presence of a group interaction is determined not by the network’s topology, but rather by the dynamics defined on the network: In the nonlinear voter model group interactions arise from update rules defined on an otherwise dyadic network.

In the Statistical Physics treatment of interacting particles, interactions are typically assumed to be pairwise, but this does not preclude the emergence of three-body or higher-order collisions, as captured by the Ursell--Mayer cluster expansion~\cite{Huang87,Pathria}. The diagrams of this expansion resemble hyperedges in modern HO interactions literature. In the virial expansion of systems with purely pairwise interaction potentials, the second-order coefficient reflects two-body collisions, while higher-order coefficients correspond to collisions among larger sets of particles. This illustrates that many-body effects—analogous to HO interactions—can arise even within fundamentally pairwise interaction frameworks.

It is in the context of the above discussion that the central questions that motivate the present work arise. What is exactly meant or implied by a ``higher-order interaction''? Does it correspond to a genuine group interaction, encoded by an explicit three-body or higher-order potential, or can it emerge from combinations of pairwise interactions? When is a hypergraph representation necessary? Under what conditions can HO interactions be reduced to effective dyadic interactions? 

In fact, the reducibility of HO interactions to effective PW interactions has been already addressed at different levels. A first microscopic approach studies when HO interactions on a hypergraph can be rewritten in terms of PW interactions~\cite{PRE_Lambiotte_multibody,Modelling_Lambiotte}. A key result of these studies, restricted to continuous state variables of the nodes ($s \in [0,1]$), is that linear interactions on hypergraphs can be always mapped onto PW interactions on a weighted complex network. 
A second approach considers reducibility at a macroscopic level, identifying dyadic-based models that reproduce the same macroscopic equations than those derived from HO interactions on a hypergraph~\cite{Sandro}.
Related to this macroscopic reducibility, is the question of the minimal model capable of reproducing macroscopic data with given precision~\cite{Peixoto2021,Malizia2024} or that of dimensionality reduction~\cite{Helcio, functionalreducibility}.
More recently, the comparison between graph-based and hypergraph-based representations of HO interactions has been discussed~\cite{Peixoto2026}.

In this work we examine the reducibility of HO interactions in \textit{social impact models}~\cite{Latane1981,Tobias2025}, defined on hypergraphs. In these models, each agent is endowed with a binary state $s=\pm 1$, and the interaction with a group of agents, represented by a hyperedge, is characterized by a social impact function. We focus on node-update dynamics, following the common modeling approach of HO interactions~\cite{ Moreno2020, Modelling_Lambiotte, Arruda2024, Iacopini2019, Sandro, Min2025}, where a focal node updates its state according to the influence exerted by one of its interaction groups. This is a genuinely HO interaction since each update depends on all agents within the hyperedge.
Our main result is that a general social impact model on an arbitrary hypergraph can be always reduced at a microscopic level to a model of PW interactions on a projected complex network with weighted links. This mapping is exact: the stochastic dynamics of HO interactions on the hypergraph is equivalent to a PW dynamics on the projected network, with link weights encoding the probability of interaction between node pairs and, in general, depending on the instantaneous state configuration.

As an illustrative class of social impact models, we introduce \textit{hypergraph-voter models}~\cite{Min2025}, in which agents update their state according to the power $q$ of the fraction of agents in the same hyperedge holding the opposite opinion. For the \textit{hypergraph-linear voter model} ($q=1$), the mapping yields a projected network with time-independent weights, which gives a transparent interpretation of the reduced dynamics. At the macroscopic level, the \textit{hypergraph-linear voter model} features the same evolution equations as the standard voter model on the projected network with unweighted links (VM)~\cite{CliffordSudbury1973, HolleyLiggett1975, Liggett1985, Liggett1999, Suchecki_2004,Suchecki_2005,Sood_2005, Vazquez_PA, Castellano2009}, implying that macroscopic behavior can be captured by assuming homogeneous weights. By contrast, the \textit{hypergraph-nonlinear voter model} ($q \neq 1$) generates effective pairwise interactions on a projected network whose weights depend on the instantaneous system state configuration. More generally—but less precisely—the nonlinear voter model on the unweighted projected network (NLVM) ~\cite{Ramirez2024,Tobias2025,Llabres2025} provides a good approximation of the ordering dynamics of hypergraph-nonlinear voter models.

The paper is organized as follows. In Sec.~\ref{sec:NU_HG}, we review the class of hypergraphs used to encode HO interactions, introduce the notion of projected networks, and define a general class of HO dynamical processes: social impact models on hypergraphs. Sec.~\ref{sec:Reducibility} addresses the mapping from HO to PW interactions. We introduce a general weighted PW interaction model and derive the projected-network weights that yield an exact microscopic equivalence with the HO dynamics on the hypergraph. In Sec.~\ref{sec:Homogeneous}, we particularize the framework to homogeneous hypergraphs, deriving their macroscopic dynamics and the corresponding effective PW representations. In Sec.~\ref{sec:ReducibilityVMs}, we apply the general framework to hypergraph-voter models on several hypergraph structures. We conclude in Sec.~\ref{sec:conclusions} with a summary of our main results. Appendices~\ref{sec:app:HG_generation}--\ref{app:NLVM_HG} provide technical details and intermediate calculations.

\begin{figure*}[t]
 \centering
 \includegraphics[width=0.49\textwidth]{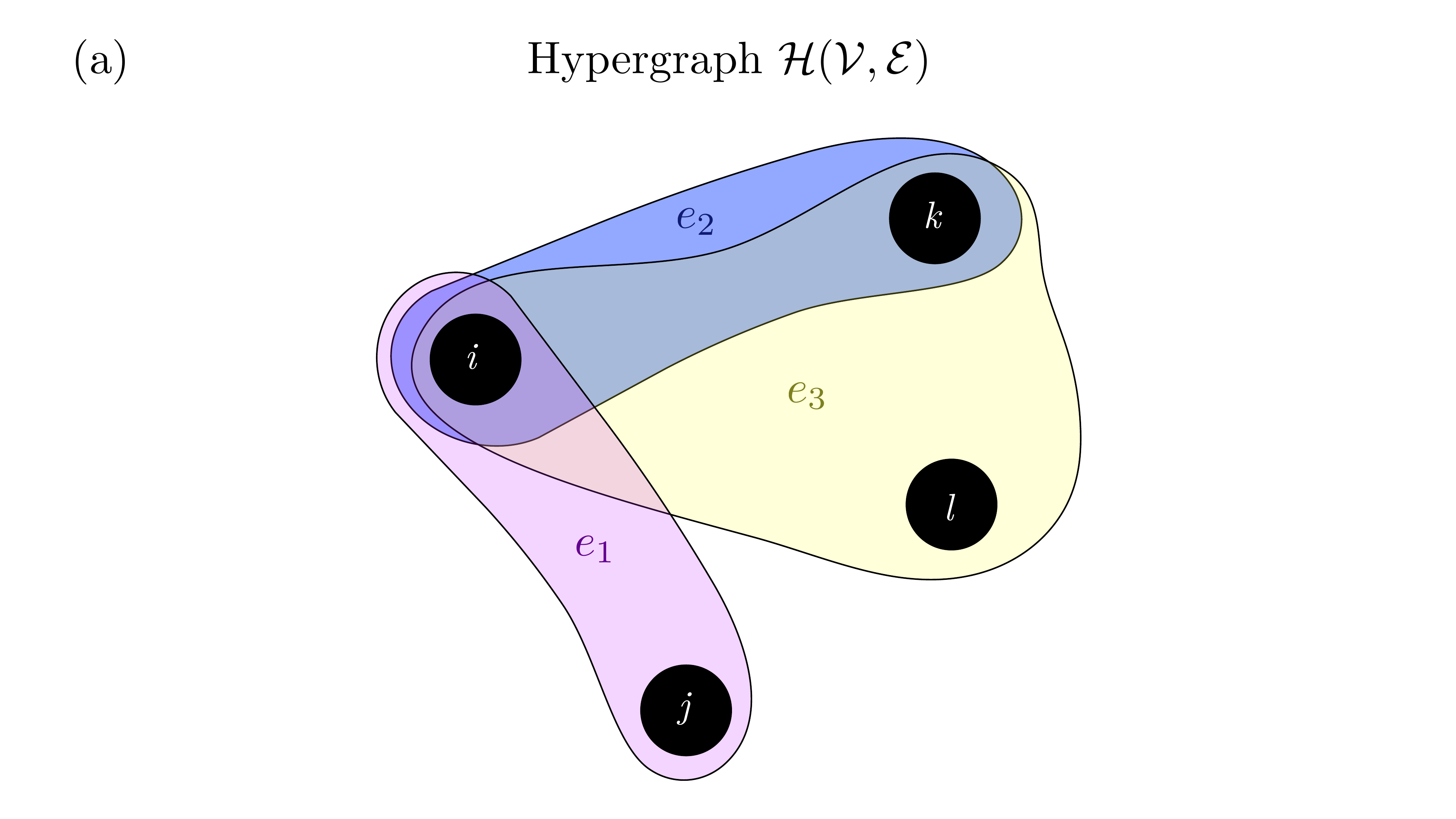}
 \includegraphics[width=0.49\textwidth]{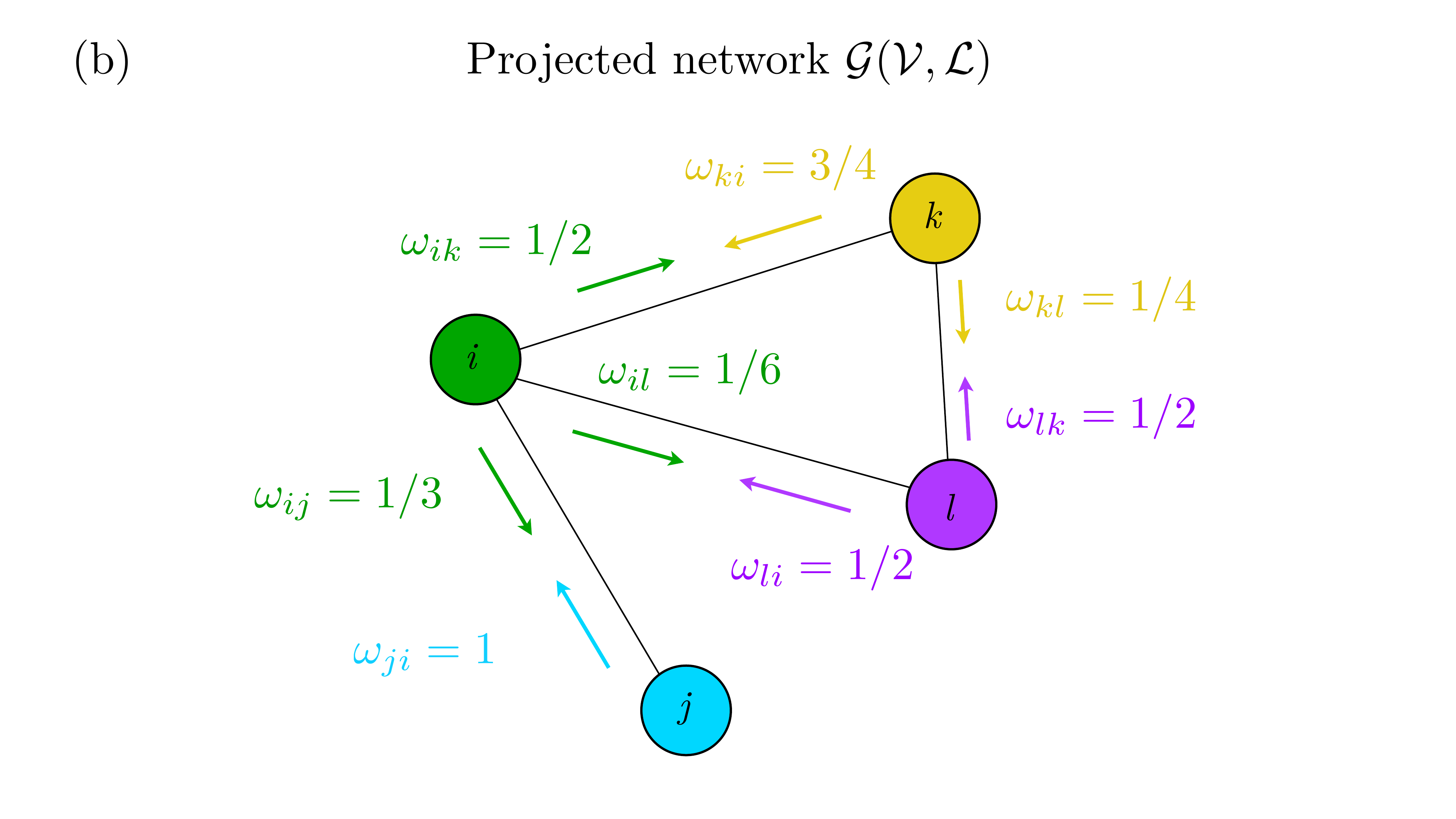}
 \caption{Illustration of (a) a hypergraph $\mathcal{H(V,E)}$ and (b) its projected network $\mathcal{G(V,L)}$. (a) The hypergraph is composed by the set $\Vc=\{i,j,k,l\}$ of $N=4$ nodes and the set $\mathcal{E}=\{e_1,e_2,e_3\}$ of $\mathcal{M}=3$ hyperedges, where $e_1=\{i,j\}$, $e_2=\{i,k\}$ and $e_3=\{i,k,l\}$ are the hyperedges with respective orders $d_{e_1}=1$, $d_{e_2}=1$ and $d_{e_3}=2$. The hyperdegrees are $\kappa_i=3$, $\kappa_j=1$, $\kappa_k=2$, and $\kappa_l=1$ and the hyperedges incident to each node read $\mathcal{E}_i=\{e_1,e_2,e_3\}$, $\mathcal{E}_j=e_1$, $\mathcal{E}_k=\{e_2,e_3\}$, and $\mathcal{E}_l=e_3$, respectively. (b) The projected network is formed by the same set of nodes $\Vc$, with degrees $k_i=3$, $k_j=1$, $k_k=2$ and $k_l=2$, and the set of links $\mathcal{L}=\bigl\{\{i,j\},\{i,k\},\{i,l\},\{k,l\}\bigr\}$. As an example, we assign the weights corresponding to the equivalent pairwise dynamics of the hypergraph-linear voter model defined in Sec.~\ref{sec:Map_VM}, Eq.~\eqref{eq:wij_VM}.}
 \label{fig:hypergraph_setup}
\end{figure*}

\section{Social Impact Models on hypergraphs}
\label{sec:NU_HG}

In this section, we introduce a general framework for the study of social impact models on hypergraphs. We first describe the topology of interactions defined by a hypergraph (Sec.~\ref{sec:hypergraph}), and then detail the general update rules governing the dynamics (Sec.~\ref{sec:dynamics}).

\subsection{Definition of hypergraphs and the projected networks}
\label{sec:hypergraph}

\textit{Hypergraphs} are the most widely used representation of higher-order interactions. A hypergraph $\mathcal{H(V,E)}$ is defined as a set $\Vc=\{1,2,..,N\}$ of $N$ nodes and a set $\mathcal{E}=\{e_1,e_2,..,e_{\mathcal M}\}$ of $\mathcal{M}$ hyperedges, where a hyperedge of order $d$ (or $d$-hyperedge) connects a subset of $d+1$ distinct nodes ($d \ge 1$). Under this convention, $1$-hyperedges correspond to standard links that connect pairs of nodes (as in ordinary complex networks), $2$-hyperedges represent triangles connecting three nodes, and higher-order hyperedges generalize this pattern. For any given node $i$, we denote by $ \Ec_i = \{e_i^{(1)},\dots,e_i^{(\kappa_i)}\} $ 
the set of hyperedges incident to $i$, that is, the collection of hyperedges containing node $i$. The hyperdegree of node $i$, denoted by $\kappa_i$, is the number of hyperedges to which it belongs.

In Fig.~\ref{fig:hypergraph_setup}(a), we show a simple example of a hypergraph of $N=4$ nodes and $\mathcal{M}=3$ hyperedges, where we describe the nodes in each hyperedge, its order, as well as the hyperedges incident on each node and its associated hyperdegree.

Although in Sec.~\ref{sec:Reducibility} we develop the mapping of HO to PW interactions for arbitrary hypergraphs, we now define several $d$-hypergraphs, i.e. those containing exclusively $d$-hyperedges, later considered in some explicit calculations. 
\begin{itemize}
 \item \textbf{Complex network:} Only $1$-hyperedges, i.e. links, exist. 
 In this case, the set of hyperedges $\mathcal{E}_i$ incident to a node $i$ is in one-to-one correspondence with the set $\Vc_i$ of its $k_i$ neighbors, where the node's hyperdegree $\kappa_i$ becomes its degree $k_i$. The hypergraph is equivalent to a standard complex network.
\item \textbf{Homogeneous hypergraphs:} Nodes are topologically equivalent.
    \begin{itemize}
     \item \textbf{Fully connected hypergraph}: A single \\ $(N{-}1)$-hyperedge connecting all $N$ nodes, i.e. $\kappa_i=1$, $\forall i=1,..,N$.
    
     \item \textbf{Complete $d$-hypergraph:} Also called the \textit{annealed $(d{+}1)$-uniform hypergraph}~\cite{Min2025}. It contains all possible $d$-hyperedges. All nodes have hyperdegree
     \begin{eqnarray}\label{eq:Md}
     M_d \equiv \binom{N-1}{d},
     \end{eqnarray}
     and the total number of hyperedges is $\binom{N}{d+1}$.
     
     This generalizes the standard complete graph, i.e., the complete $1$-hypergraph, in which each node is linked to any other node, to higher-order interactions. Unlike the fully connected hypergraph where each node is connected to all other $N-1$ nodes by the same hyperedge, in the complete $d$-hypergraph each node is connected to any other node by $\binom{N-2}{d-1}$ $d$-hyperedges. In the limit $d=N-1$, the complete $d$-hypergraph reduces to the fully connected hypergraph discussed above.
    \end{itemize}
\item \textbf{Random hypergraphs:}
    \begin{itemize}
     \item \textbf{Erd\H{o}s–R\'enyi (ER) $d$-hypergraph:} This the generalization of ER random graphs~\cite{e1959,r1959, er1960} to $d$-hypergraphs~\cite{Bollobas_Erdos_1976, bookHG}, where all hyperedges are of order $d$. With probability $p$, each of the $\binom{N}{d+1}$ possible subsets of $d{+}1$ nodes is included as a hyperedge. As a result, the hyperdegree distribution $P_d(\kappa)$ is the fraction of nodes with $\kappa$ incident $d$-hyperedges, randomly chosen from a total of $M_d$ possible incident hyperedges, defined in Eq.~(\ref{eq:Md}). It follows a binomial law with single-event probability $p$, namely $P_d(\kappa) = \textbf{B}(\kappa,M_d;p)$ where
     \begin{equation} \label{eq:binomial}
     \textbf{B}(n,d;x)\equiv\binom{d}{n} x^{n}(1-x)^{d-n},
     \end{equation}
     is the general binomial distribution.
     
     The expected number of hyperedges $E_d$ and the average hyperdegree $\mu_d$ are
     \begin{align}
     E_d &= p \binom{N}{d+1}, \\
     \mu_d &= p M_d= (d+1) \frac{E_d}{N}.
     \end{align} 
     ER $1$-hypergraphs correspond to ER networks. An ER $d$-hypergraph becomes a complete $d$-hypergraph when $p=1$, and a fully connected hypergraph when, additionaly, $d=N-1$.
     \item \textbf{$z$-regular random (RR) $d$-hypergraph:} Each node has exactly $z$ incident $d$-hyperedges, chosen at random among the possible $M_d$ incident hyperedges. In this way, the hyperdegree distribution is given by the Kronecker-delta function $P_d(\kappa)=\delta_{\kappa,z}$ and the total number of hyperedges is fixed $E_d=N z / (d+1)$. $z$-RR $1$-hypergraphs correspond to $z$-RR networks~\cite{Bollobas_2001}.
    \end{itemize}
Details on the generation of ER $d$-hypergraphs are given in Appendix~\ref{sec:app:HG_generation}. It is worth noting that the generation of both hypergraphs may occasionally produce disconnected topologies. Throughout this work, we focus exclusively on connected hypergraphs when implementing the dynamics.
\end{itemize}

\subsubsection{Projected network}
\label{sec:projected_net}
Besides the hypergraph representation, we also introduce the concept of a \textit{projected network}. Given a hypergraph $\mathcal{H(V,E)}$, one can construct its projected network $\mathcal{G(V,L)}$ on the same set of nodes $\Vc$ by connecting with a link each pair of nodes that share at least one hyperedge, a simple operation called nominal projection~\cite{Proj_ntwrk}.

We denote the degree distribution of the projected network as $P(k)$ and its average value as $\mu \equiv \langle k \rangle$. An illustration is given in Fig.~\ref{fig:hypergraph_setup}(b), which shows the projected network of a hypergraph with $N=4$ nodes and $\mathcal{M}=3$ hyperedges. 

We define as \textit{overlap} the situation in which two nodes share more than one hyperedge. Details on the characterization of projected networks and overlap for ER and $z$-RR $d$-hypergraphs are given in Appendix~\ref{sec:app:Mapping_HG} and Appendix~\ref{app:Overlap}, respectively. In sparse $d$-hypergraphs, where overlap is negligible, the degree $k_i$ of a node $i$ in the projected network relates to its hyperdegree $\kappa_i$ in the hypergraph as
\begin{equation}
k_i \simeq d \, \kappa_i.
\end{equation}
Additionally, both the fully connected hypergraph and the complete $d$-hypergraph project onto the complete graph, although in the latter case there is a high degree of overlap as each pair of nodes belongs to $\binom{N-2}{{d-1}}$ distinct hyperedges.

The projected network representation has the advantage of preserving the neighborhood structure of each node, as well as encoding its pairwise interactions with neighbors by means of a simpler topology. However, it also has limitations: it loses explicit information about the higher-order interactions of the original hypergraph. In particular, overlap cannot be captured by the projected network.

\subsubsection{Weighted projected network}
\label{sec:weighted projected_net}

The above discussion motivates the definition of the \textit{weighted projected network}~\cite{Proj_ntwrk, Carletti2020}. If the pair of nodes $i$ and $j$ belong to $m$ common hyperedges, one assigns a weight $ \oij = m$, to the link $(i,j)$, thereby encoding the contribution of these shared hyperedges and quantifying the effective pairwise interaction between $i$ and $j$. In this way, the weighted projected network encodes specific information about the topology of higher-order interactions.

More generally, one can define alternative projection rules where the weights $\oij$ specify how higher-order interactions are translated into effective pairwise interactions. In particular, the weights can be chosen to depend on the \emph{dynamics} under consideration, so that the microscopic transition probabilities of the original model on the hypergraph are exactly reproduced by an equivalent dynamics on the weighted projected network, even though the dynamical rules may differ. As we shall see in Sec.~\ref{sec:Reducibility}, the precise definition of the weights ensuring such equivalence depends on both the hypergraph structure and the specific dynamical process. Throughout this work, the term weighted projected network will refer to this latter case, where the weights are dynamically defined rather than simply given by the number of shared hyperedges.

\subsection{Dynamical model}
\label{sec:dynamics}
We introduce social impact models on hypergraphs, which constitute the basic dynamical framework of our analysis. These models describe HO interactions among nodes belonging to the same hyperedge. We consider a population of $N$ nodes, each endowed with a binary state variable $s_i \in \{-1,+1\}$, with $i=1,\dots,N$. The full microscopic configuration is denoted by $\mathbf{S} \equiv (s_1,\dots,s_N)$.

The dynamics follows a node-update process defined as follows. At each time step $\Delta t = 1/N$:
\begin{enumerate}
 \item Select a node $i$ at random. Let $s_i$ be its state and $\kappa_i$ its hyperdegree.
 \item Select randomly one of the $\kappa_i$ incident hyperedges $e\in\Ec_i$. Let $d_e$ be its order and $n_e$ the number of nodes, excluding $i$, with state $+1$ that it contains. 
 \item Node $i$ flips its state, i.e. $s_i \to -s_i$, with a probability $f(\phi_e)$, where $\phi_e$ is the fraction of nodes (excluding node $i$) in the hyperedge $e$ that are in the opposite state to $s_i$, 
 \begin{equation} \label{eq:phi_ei}
 \phi_e=
 \begin{cases}
 \dfrac{n_e}{d_e}& \text{if $s_i=-1$,} \\
 1-\dfrac{n_e}{d_e}& \text{if $s_i=+1$.} 
 \end{cases}
 \end{equation}
\end{enumerate} 
Time is measured in Monte Carlo steps (MCS). One MCS corresponds to $N$ updating attempts such that, on average, each node is selected once per MCS.

This dynamical model adopts a node-update \emph{passive communication} perspective, where a focal node receives input from a group of nodes that belong to an incident hyperedge and updates its state according to their states. 

The function $f(\phi)$ is known as the \textit{social impact function} introduced in social psychology in Ref.~\cite{Latane1981}. In the context of opinion dynamics, this function specifies how the persuasive power of a group scales with its relative size. For any function satisfying $f(0)=0$, the system presents two absorbing states, corresponding to a situation in which all nodes are in the same state. The step function $f(\phi)=\Theta(\phi-T)$ corresponds to the case of the threshold model \cite{Granovetter,Watts}, in which a node flips state only if a fraction $\phi \ge T$ of its neighbors are in the opposite state, where $T \in [0,1]$ is the threshold. A particular form of complex contagion \cite{Centola2018} corresponds to the case $T=1$, where a node flips when all its neighbors are unanimously opposite \cite{Iacopini2019,Moreno2020}. 
Recent studies show that the specific form of the social impact function governs the macroscopic ordering dynamics~\cite{Tobias2025}.

Summing over all hyperedges incident to node $i$, the updating rules yield the microscopic transition probability:
\begin{equation} \label{eq:wi_HG}
 \Omega(s_i\to -s_i) = \frac{1}{\kappa_i}\sum_{e\in \Ec_i} f\!\left(\phi_e\right),
\end{equation}
where the term $f(\phi_e)/\kappa_i$ accounts for the contribution of hyperedge $e$.

When all hyperedges are of order $d=1$, the fraction $\phi_e$ can only take the values $0$ or $1$. Consequently, whenever $f(0)=0$ and $f(1)=1$, the social impact dynamics on complex networks reduces to the VM, where an agent selected at random adopts the state of a randomly chosen neighbor~\cite{CliffordSudbury1973, HolleyLiggett1975, Liggett1985, Liggett1999, Suchecki_2004,Suchecki_2005,Sood_2005, Vazquez_PA, Castellano2009}. In this case, the flipping probability becomes
\begin{equation} \label{eq:wi_VM}
 \Omega(s_i\to -s_i) = \frac{1}{k_i} \sum_{j \in \Vc_i} \delta_{s_j,-s_i} = \frac{k_{i,-s_i}}{k_i},
\end{equation}
where $k_{i,-s_i}$ is the number of neighbors of $i$ in the opposite state. 

\section{Reducibility of social impact models on hypergraphs to pairwise interactions}
\label{sec:Reducibility}

In this section, we examine how the social impact models on hypergraphs, introduced in Sec.~\ref{sec:dynamics}, can be mapped at the microscopic level onto equivalent \textit{effective pairwise interactions} on a complex network. Our goal is to develop a method that exactly represents HO interactions in terms of equivalent PW ones.

\subsection{Model for pairwise interactions} \label{sec:general_pw}

A natural reference point for a reduction to PW dynamics is the VM~\cite{CliffordSudbury1973, HolleyLiggett1975, Liggett1985, Liggett1999, Suchecki_2004,Suchecki_2005,Sood_2005, Vazquez_PA, Castellano2009}, which stands as the paradigmatic example of pairwise imitation dynamics. As stated before, in its standard form, a focal node copies the state of one of its neighbors selected at random, providing the minimal framework for studying PW interactions. 

Here we consider a general PW imitation model in which each node $i$ is connected to a set of neighbors $\Vc_i$, and its degree $k_i$ is defined as its total number of neighbors. The topology and directionality of interactions are encoded in a weighted adjacency matrix $\{\oij\}$, such that $\oij$ denotes the probability that node $i$ selects a neighbor $j$ to copy its state in a single time step. In this way, $\oij=0$ if $j$ is not a neighbor of $i$ ($j \not \in \Vc_i$). In general, the weights need not be symmetric, $\oij \ne \omega_{ji}$, and they are normalized for each node,
\begin{equation}
 \sum_{j \in \Vc_i} \oij = 1,
 \label{eq:normalization}
\end{equation}
ensuring that node $i$ always selects one of its neighbors during an update. In general, we include the possibility that the weights might depend on the current state configuration $\mathbf{S}$. Therefore, the weights $\{\oij\}$ are not only a topological property, but they are part of the definition of the dynamics.

The model is governed by the following updating rules. At each time step $\Delta t=1/N:$
\begin{enumerate}
 \item Select a node $i$ at random.
 \item Select one of its neighbors $j$ with probability $\oij$.
 \item Node $i$ copies the state of node $j$, $s_i\to s_j$.
\end{enumerate}

These updating rules lead to the following flipping probability for node $i$:
\begin{equation} \label{eq:omega-flip}
 \Omega(s_i \to -s_i)= \sum_{j\in \Vc_i} \oij \, \delta_{s_j,-s_i}.
\end{equation}
Note that this model presents two absorbing states, where all nodes are in the same state, either $+1$ or $-1$. 

According to the rules above, the update of node $i$ involves the selection of the neighbor $j$ (step 2), and the pairwise interaction with that neighbor (step 3). While the choice of $j$ is determined by $\oij$, the interaction depends only on the states $s_i$ and $s_j$ and not on the states of other nodes. It is in this sense that we consider that the interactions are pairwise.

For \emph{degree-homogeneous weights}, that is, $\omega_{ij}=1/k_i$ for all $j\in\Vc_i$, a neighbor is selected at random and the dynamics reduces to that of the standard VM~\cite{Vazquez_PA,Suchecki_2004,Suchecki_2005, Sood_2005, Castellano2009}, with flipping probabilities given by Eq.~\eqref{eq:wi_VM}. Note that in this case the choice of $j$ does not depend on the states $s_i$ and $s_j$.

Another relevant case is that of the \emph{complete graph with state-homogeneous weights}, where all nodes in the same state are statistically equivalent. Consequently, the elements of the adjacency matrix depend only on the states of the nodes, $\omega_{ij}=\omega(s_i,s_j)$ for all $i,j$, allowing for four possible values, denoted by $\omega(\pm,\pm)$. Applying the normalization condition in Eq.~\eqref{eq:normalization} to nodes in states $s_i=+1$ and $s_i=-1$, respectively, yields
 \begin{subequations}
 \begin{align}
 (N_+-1)\,\omega(+,+)+N_-\,\omega(+,-)&=1,\\
 N_+\,\omega(-,+)+(N_--1)\,\omega(-,-)&=1,
 \end{align}
 \label{eq:normalization-CG}
 \end{subequations}
where $N_\pm$ denote the number of nodes in state $\pm1$, and we define the density of nodes in state $+1$ as $x \equiv N_+/N$. This result implies that the weights $\omega(\pm,\pm)$ are not independent. In the case of uniform weights $\omega(\mp,\pm)=1/(N-1)$, the classical VM on the complete graph~\cite{VM_CG} is recovered. 
 
The weights $\omega(\pm,\pm)$ allow us to compute the flipping probabilities from Eq.~\eqref{eq:omega-flip} as
\begin{subequations}
 \begin{align}
 \Omega(- \to +) &= \omega(-,+) Nx, \\
 \Omega(+ \to -) &= \omega(+,-) N(1-x).
 \end{align}
 \label{eq:omega_PW_CG}
\end{subequations}

\subsection{Reducibility of social impact models on hypergraphs}
\label{sec:reduc-HG}

In this section, we develop a mapping of the social impact model with HO interactions on a hypergraph, defined in Sec.~\ref{sec:dynamics}, onto an equivalent model with PW interactions, introduced in Sec.~\ref{sec:general_pw}. 

At each time step, we require the microscopic transition probabilities of the HO dynamics on the hypergraph, Eq.\eqref{eq:wi_HG}, to coincide with those of the PW dynamics on the projected network, Eq.~\eqref{eq:omega-flip}. This condition yields a set of constraints on the weighted adjacency matrix $\{\oij\}$ encoding the effective pairwise interactions. We decompose each weight $\omega_{ij}$ into contributions $\omega_{ij}^e$ associated with hyperedges $e \in \mathcal{E}_i$ that contain $j$,
\begin{equation}
 \omega_{ij} = \sum_{e \in \mathcal{E}_i \mid j \in e} \omega_{ij}^e.
 \label{eq:wij}
\end{equation}

The weights $\omega_{ij}^e$ are determined by imposing the following two conditions:
\begin{enumerate}
 \item Probability conservation per hyperedge: The probability that a node $i$ randomly selects an incident hyperedge $e$ in the HO dynamics must be equal to the probability that $i$ copies one of its neighbors $j \in e$ in the PW interaction, 
 \begin{align}
 \frac{1}{\kappa_i} =\sum_{j \in e} \oij^e,
 \label{eq:map_ei}
 \end{align}
 where we recall that $\kappa_i $ is the hyperdegree of node $i$. This condition ensures that a neighbor $j$ is always selected, $\sum_{e\in \Ec_i} \sum_{j \in e} \oij^e = 1$.
 \item Flip consistency: Given the selected hyperedge $e$, the flipping probability of node $i$ in the HO dynamics must be equal to the probability that $i$ flips its state by copying a neighbor $j \in e$ with opposite state $s_j=-s_i$. This leads to the condition:
 \begin{align}
 \frac{1}{\kappa_i} f\left(\phi_{e} \right)=\sum_{j \in e} \oij^e \, \delta_{s_j,-s_i},
 \label{eq:map_ei_flip}
 \end{align}
 where we recall that $\phi_{e} $ is the fraction of nodes in the opposite state to $i$ in $e$, given by Eq.~\eqref{eq:phi_ei}.
\end{enumerate} 
Equations~\eqref{eq:map_ei} and \eqref{eq:map_ei_flip} imply that $f(0)=0$ and $f(1)=~1$. Therefore, the present mapping is restricted to social impact functions with absorbing states. Together, these equations ensure that the PW dynamics reproduces the microscopic transition probabilities of the HO interaction model. Equation~\eqref{eq:map_ei} fixes the total probability assigned to the neighbors of $i$ within each hyperedge $e$, while Eq.~\eqref{eq:map_ei_flip} constrains how that probability is distributed between neighbors in opposite and equal states. However, these relations determine only the total probability assigned to each subset of neighbors in $e$. As a result, many distinct sets of values $\{\oij^e\}$ satisfy the required conditions in Eqs.~\eqref{eq:map_ei} and \eqref{eq:map_ei_flip}. To resolve this indeterminacy, we impose the natural condition that all neighbors $j\in e$ sharing the same state are statistically equivalent within $e$ and therefore receive the same weight, i.e. $\omega^e_{ij}=\omega_i^e(s_i,s_j)$. This condition, together with Eq.~\eqref{eq:map_ei_flip}, leads to the following weights for links connecting nodes in opposite state:
\begin{equation} \label{eq:wij_ei_flip}
 \omega^e_i(s_i,-s_i) =
 \begin{cases}
 \dfrac{1}{\kappa_i n_e}f\left(\dfrac{n_e}{d_e}\right), & \text{if}~s_i=-1, \\\\
 \dfrac{1}{\kappa_i (d_e - n_e)}f\left(1-\dfrac{n_e}{d_e}\right), & \text{if}~s_i=+1,
 \end{cases}
\end{equation}
where we recall that $d_e$ is the order of hyperedge $e$, i.e. the number of nodes excluding $i$ in $e$, of which $n_e$ are in state $+1$. We emphasize that the weights $\omega_i^e(s_i,-s_i)$ are defined for neighbors $j\in e$ such that $s_j=-s_i$. Accordingly, $ n_e\in~[1,\,d_e]$ if $s_i=-1$, while $n_e \in~[0,\, d_e-1]$ if $s_i=+1$, and the corresponding denominators never vanish.

The other two weights are obtained from Eqs.~\eqref{eq:map_ei} as
\begin{equation} \label{eq:wij_ei}
 \omega^e_i(s_i,s_i) =
 \begin{cases}
 \dfrac{1}{\kappa_i (\de -n_e)}\left[1-f\left(\dfrac{n_e}{d_e}\right)\right], & \text{if } s_i=-1,\\
 \\
 \dfrac{1}{\kappa_i n_e}\left[1-f\left(1-\dfrac{n_e}{d_e}\right)\right], & \text{if } s_i=+1,
 \end{cases}
\end{equation}
where, again, the denominators are nonzero by construction, since the weights are defined only for neighbors sharing the same state.

Finally, using Eq.~\eqref{eq:wij}, the total weight $\oij$ is obtained as the sum of the contributions $\omega_i^e(s_i,s_j)$ over all hyperedges $e$ containing both nodes, providing an exact mapping for any hypergraph and social impact function $f$. We highlight that each weight $\omega_{i}^e(s_i,s_j)$ depends on the local state configuration through $\phi_e$, defined in Eq.~\eqref{eq:phi_ei}, and therefore the weighted adjacency matrix is, in general, implicitly time-dependent.

\section{Social impact models on homogeneous hypergraphs} \label{sec:Homogeneous}

While in Sec.~\ref{sec:Reducibility} we establish an exact microscopic mapping of social impact models on arbitrary hypergraphs onto PW dynamics, here we focus on homogeneous hypergraphs, for which nodes sharing the same state are statistically equivalent.

We first derive the macroscopic evolution equations for this general class of hypergraphs. We then particularize the analysis to the fully connected hypergraph and to complete $d$-hypergraphs by explicitly computing the corresponding transition probabilities. Finally, we obtain explicit expressions for the effective weights, given in Eqs.~\eqref{eq:wij_ei_flip} and~\eqref{eq:wij_ei}, which define a projected network that reduces, in each case, to a complete graph with state-homogeneous weights.

\subsection{Macroscopic dynamics}
The macroscopic dynamics is described by the time evolution of the global density $x$ of nodes in state $+1$. This quantity is a stochastic variable that changes by $\pm 1/N$ in a single update, with transition probabilities $T^\pm(x) \equiv T(x \to x\pm 1/N)$ given by
\begin{align}\label{eq:T_global}
 T^\pm(x) & = \sum_{i|s_i=\mp1}\,\frac{1}{N} \,\Omega(s_i\to -s_i),
\end{align}
which in turn determines the evolution of the ensemble average,
\begin{equation}
\frac{d\langle x\rangle}{dt}=\left\langle T^+(x)-T^-(x)\right\rangle.
\label{eq:dxdt_general}
\end{equation}
In general, the probabilities $T^\pm(x)$ depend on the full microscopic configuration $\mathbf{S}$, making the average in Eq.~\eqref{eq:dxdt_general} intractable. However, in homogeneous hypergraphs, nodes sharing the same state are statistically equivalent, implying that the transition probabilities $\Omega(\pm \to \mp)$ depend only on the state of the focal node and on the global density $x$, rather than on the detailed configuration. As a result, the global transition probabilities simplify to
\begin{subequations} \label{eq:T_global-1}
 \begin{align}
 T^+(x) & = (1-x)\,\Omega(- \to +), \\
 T^-(x) & = x\,\Omega(+ \to -).
 \end{align}
\end{subequations}
Furthermore, when focusing on the thermodynamic limit $N\to\infty$, finite-size fluctuations vanish and ensemble averages can be replaced by their deterministic values. The dynamical equation then becomes closed in the single variable $x$,
\begin{equation}
\frac{dx}{dt}= (1-x)\,\Omega(- \to +) - x\,\Omega(+ \to -),
\label{dxdt_Omega}
\end{equation}
which provides a direct way to determine the fixed points of the macroscopic dynamics and assess their stability.

\subsection{Fully connected hypergraph} \label{sec:wpn_fully}
As explained in Sec.~\ref{sec:projected_net}, the fully connected hypergraph consists of a single $(N-1)$-hyperedge, so that $\kappa_i=1$ for all $i$. The transition probabilities of the HO dynamics, obtained from Eq.~\eqref{eq:wi_HG}, therefore read
\begin{subequations} 
\begin{align}
 \Omega(- \to +) &=f\left(\frac{N_+}{N-1}\right)\simeq f(x), \\
 \Omega(+ \to -) &=f\left(\frac{N_-}{N-1}\right)\simeq f(1-x),
\end{align}
\label{eq:Omega-FCHG}
\end{subequations}
where the approximation holds in the $N \gg 1$ limit. 
Together with Eq.~\eqref{dxdt_Omega}, these expressions provide a complete description of the macroscopic dynamics in terms of the collective variable $x$.

The HO dynamics defined on the fully connected hypergraph is mapped onto a PW dynamics on the complete graph with state-homogeneous weights that can be obtained from Eqs.~\eqref{eq:wij_ei_flip} and \eqref{eq:wij_ei} as
\begin{subequations}
 \begin{align}
 \label{eq:omega-+}
 \omega(-,+) &= \frac{1}{N_+} f \left(\frac{N_+}{N-1} \right) \simeq \frac{f(x)}{Nx}, \\
 \label{eq:omega--}
 \omega(-,-) &= \frac{1}{N_{-}-1} \left[ 1- f \left(\frac{N_+}{N-1} \right) \right] \simeq \frac{1-f(x)}{N(1-x)}, \\
 \label{eq:omega+-}
 \omega(+,-) &= \frac{1}{N_{-}-1} f \left(\frac{N_{-}-1}{N-1} \right) \simeq \frac{f(1-x)}{N(1-x)}, \\
 \label{eq:omega++}
 \omega(+,+) &= \frac{1}{N_+} \left[1 - f \left(\frac{N_{-}-1}{N-1} \right) \right] \simeq \frac{1-f(1-x)}{Nx}.
 \end{align} 
 \label{eq:weights-FCH}
\end{subequations}

For consistency, we note that these weights can also be obtained by equalizing the flipping probabilities of the HO dynamics on the fully connected hypergraph, Eqs.~\eqref{eq:Omega-FCHG}, with those of the PW dynamics on the complete graph with state-homogeneous weights, Eq.~\eqref{eq:omega_PW_CG}, together with the normalization condition, Eq.~\eqref{eq:normalization-CG}. 

\subsection{Complete $d$-hypergraph}
\label{sec:wpn_CdHG}

For the complete $d$-hypergraph, each node belongs to all possible $d$-hyperedges, so that $\kappa_i=M_d=\binom{N-1}{d}$ for all $i$, as defined in Eq.~\eqref{eq:Md}. In this case, the sum over incident hyperedges in Eq.~\eqref{eq:wi_HG} can be reorganized as a sum over the possible compositions of a $d$-hyperedge, yielding the transition probabilities
\begin{subequations} 
 \begin{align}
 \Omega(- \to +)&= \sum_{n=0}^d \textbf{H}(N_+, N_--1,d, n) f\left(\frac{n}{d}\right), \\
 \Omega(+ \to -)&=\sum_{n=0}^{d} \textbf{H}(N_+-1, N_-, d, n) f\left(1-\frac{n}{d}\right),
 \end{align}
 \label{eq:omega_CdH}
\end{subequations}
where $\textbf{H}(N_+, N_-, d, n)$ is the hypergeometric distribution defined as 
\begin{equation}\label{eq:hyper_def}
 \textbf{H}(N_+, N_-, d, n)=\frac{\binom{N_+}{n} \binom{N_{-}}{d-n}}{\binom{N_++N_-}{d}}.
\end{equation}

The macroscopic description of the system is fully described by Eqs.~\eqref{dxdt_Omega} and~\eqref{eq:omega_CdH}.

To compute the weights, Eqs.~\eqref{eq:wij_ei_flip} and \eqref{eq:wij_ei}, of the equivalent PW dynamics defined on the projected network (complete graph), we note that any pair of nodes $\{i,j\}$ is contained in exactly $\binom{N-2}{d-1}$ hyperedges. However, the contribution $\omega_i^e(s_i,s_j)$ of a given hyperedge $e$ depends on its local state configuration. The weight $\omega(-,+)$ is computed by fixing $s_i=-1$ and $s_j=+1$. Let $n\ge1$ denote the total number of nodes in state $+1$ within a hyperedge $e$, including node $j$. There are $\binom{N_+-1}{n-1}\binom{N_--1}{d-n}$ hyperedges with a given value of $n$, all contributing the same amount. Using Eqs.~\eqref{eq:wij} and \eqref{eq:wij_ei_flip}, the total weight $\omega(-,+)$ therefore reads
\begin{equation}
\label{eq:omega-+CdHG}
 \omega(-,+)
 = \sum_{n=1}^d
 \frac{1}{M_d\, n}
 \binom{N_+ - 1}{n-1}\binom{N_- - 1}{d-n}
 f\!\left(\frac{n}{d}\right),
\end{equation}
where the sum over hyperedges containing the pair $\{i,j\}$ has been reorganized as a sum over hyperedges with identical state compositions.

Equation~\eqref{eq:omega-+CdHG} can be rewritten as
\begin{align}
 \omega(-,+) &= \frac{1}{N_+} \sum_{n=1}^d\textbf{H}(N_+, N_--1, d, n) f\left(\frac{n}{d}\right).
 \label{eq:omega-+CdHG-2}
\end{align}

As shown in Appendix~\ref{sec:app:hypergeometric-binomial}, in the limit $N \gg d$ the hypergeometric distribution  $\textbf{H}(N_+, N_-, d, n)$ can be approximated by the binomial distribution $\textbf{B}(n,d;x)$, defined in Eq.~\eqref{eq:binomial}. Using this approximation, and proceeding analogously for the remaining cases, we obtain
\begin{subequations}
 \begin{align}
 \label{eq:omega-+F}
 \omega(-,+)& \simeq \frac{F(d,x)}{Nx},\\
 \label{eq:omega--F}
 \omega(-,-) & \simeq \frac{1-F(d,x)}{N(1-x)}, \\
 \label{eq:omega+-F} 
 \omega(+,-) & \simeq \frac{F(d,1-x)}{N(1-x)}, \\
 \label{eq:omega++F}
 \omega(+,+) & \simeq \frac{1-F(d,1-x)}{Nx},
 \end{align}
 \label{eq:w_CH}
\end{subequations}
where we have defined 
\begin{equation}
 F(d,x) \equiv \left< f\left(\frac nd\right)\right>= \sum_{n=1}^d\textbf{B}(n,d,x) f\left(\frac{n}{d}\right),
 \label{eq:Fdx}
\end{equation} 
as the average of the social impact function over the binomial distribution $\textbf{B}(n,d;x)$. In the limit $d=N-1$, the function $F(d,x)$ coincides with $f(x)$, and the weights of the fully connected hypergraph, Eqs.~\eqref{eq:weights-FCH}, are recovered. 

For consistency, we note that the weights ~\eqref{eq:w_CH} can also be obtained by matching the flipping probabilities of the HO dynamics on the complete $d$-hypergraph, Eq.~\eqref{eq:omega_CdH}, with those of the PW dynamics on the complete graph with state-homogeneous weights, Eq.~\eqref{eq:omega_PW_CG}, together with the normalization condition, Eq.~\eqref{eq:normalization-CG}. 

\section{Hypergraph-Voter Models}
\label{sec:ReducibilityVMs}

In this section we consider \textit{hypergraph-voter models} defined by a special form of the 
the social impact function on our general dynamical model on hypergraphs (Sec.~\ref{sec:dynamics}),
\begin{equation}\label{eq:SIF_GVM}
f(\phi) = \phi^q,
\end{equation}
where $q>0$ is a real parameter. While hypergraph-voter models for integer $q$ have been previously studied on the complete $d$-hypergraph~\cite{Min2025}, a general formulation for $q$ real and arbitrary hypergraphs is still lacking.

The case $q=1$, which we refer to as the \emph{hypergraph-linear voter model}, constitutes the natural extension of the standard voter model on complex networks~\cite{Suchecki_2004,Suchecki_2005,Sood_2005,Vazquez_PA,Castellano2009} to hypergraphs. The case $q\neq1$, which we term the \emph{hypergraph-nonlinear voter model}, is a generalization to group interactions of arbitrary size of the NLVM~\cite{Ramirez2024,Tobias2025,Llabres2025}. 
The functional form of Eq.~\eqref{eq:SIF_GVM} satisfies $f(0)=0$, implying the existence of two absorbing consensus states, and $f(1)=1$. As a consequence, when all hyperedges are of order $d=1$, i.e., on a complex network, the dynamics coincides with that of the voter model for any value of $q$ [see Eq.~\eqref{eq:wi_VM}]. In contrast, for $d>1$, the combined effects of the nonlinear parameter $q$ and size of the group interaction $d$ lead to a genuinely different class of models.

We first analyze the macroscopic dynamics of hypergraph-voter models on homogeneous hypergraphs, and subsequently address their reducibility to effective PW interactions in general hypergraphs with detailed numerical results for Erd\H{o}s--R\'enyi (ER) and $z$-regular random (RR) $d$-hypergraphs.

\subsection{Macroscopic dynamics on homogeneous hypergraphs}
\label{sec:HGVM_dynamics}

We now derive mean-field evolution equations in the limit $N\to \infty$ for several homogeneous hypergraphs.

\begin{itemize}
 \item \textbf{Fully connected hypergraph:} Substituting the flipping probabilities given in Eqs.~\eqref{eq:Omega-FCHG} into Eq.~\eqref{dxdt_Omega}, we obtain
\begin{equation}
\frac{d x}{dt}= x(1-x) \left[ x^{q-1}-(1-x)^{q-1} \right],
\label{eq:dxdt_FCY}
\end{equation}
where we have used that in the limit $N\to\infty$ the fraction $\phi$ coincides with the global density $x$. The macroscopic dynamics is the same than the one of the NLVM on the complete graph~\cite{Ramirez2024}. 

Note that the dynamics of hypergraph-voter models on the complete graph and on the fully connected hypergraph only coincide in the linear case $q=1$.

\item \textbf{Complete $d$-hypergraph}: Substituting the flipping probabilities given in Eqs.~\eqref{eq:omega_CdH} into Eq.~\eqref{dxdt_Omega}, we obtain
\begin{equation}
 \frac{dx}{dt} = x(1-x)(2x-1)H_d(x;q).
 \label{eq:dxdt}
\end{equation}
The derivation of this equation, given in Appendix ~\ref{app:Hdq}, with the explicit form $H_d(x;q)$, relies on the fact that, in the limit $N\to\infty$, the hypergeometric distribution of Eqs.~\eqref{eq:omega_CdH} reduces to the binomial distribution, see Appendix~\ref{sec:app:hypergeometric-binomial}. 

This factorized form, in terms of the function $H_d(x;q)$, allows for a direct identification of the fixed points of the dynamics: the symmetric state at $x=1/2$ and the absorbing states at $x=0$ and $x=1$, which exist for all $d>1$ and $q\neq1$. As shown in Appendix~\ref{app:Hdq}, the function $H_d(x;q)$ does not introduce additional fixed points within the physical interval $x\in[0,1]$, but rather determines the stability of the existing ones. For $q<1$, one finds $H_d(x;q)<0$, implying that the symmetric state $x=1/2$ is stable. Conversely, for $q>1$, $H_d(x;q)>0$, and the absorbing states $x=0$ and $x=1$ become stable, thus reproducing the characteristic behavior of the NLVM on the complete graph~\cite{Ramirez2024}.

For integer values of $q$ and arbitrary $d$, the model reduces to the \emph{group-driven voter model} introduced in Ref.~\cite{Min2025}. We refer the reader to that work for a detailed discussion.

For $d=1$, $H_1(x;q)$ vanishes and the dynamics reduces to that of the standard voter model, as discussed above. For small integer values of $q$, explicit results can be obtained. In particular,
\begin{align}
 H_d(x;q=2)&= 1-\frac{1}{d}, \label{eq:Hd_q=2} \\
 H_d(x;q=3)&= 1-\frac{1}{d^2},
\end{align}
showing that hyperedges of different orders $d$ simply rescale the characteristic time scale of the macroscopic dynamics. 
For integer $q>3$, closed expressions can also be derived, although $H_d(x;q)$ then becomes an explicit function of the global density $x$. For non-integer real values of $q$, no closed-form expression exists, and $H_d(x;q)$ must be evaluated numerically. Finally, in the limit $d=N-1$ one finds
\begin{equation}
 \lim_{N\to\infty} H_{N-1}(x;q)
 = \frac{x^{\,q-1} - (1-x)^{\,q-1}}{2x-1},
\end{equation}
which, when substituted into Eq.~\eqref{eq:dxdt}, recovers the fully connected hypergraph result Eq.~\eqref{eq:dxdt_FCY}.
\end{itemize}

\subsection{Reducibility of hypergraph-voter models}

In this section, we illustrate our general framework of Sec.~\ref{sec:reduc-HG} discussing the reducibility of hypergraph-voter models to effective PW dynamics on the projected network. 

Beyond this microscopic mapping, we assess the dynamical relevance of the PW description by comparing three different dynamics: the HO dynamics on hypergraphs, the PW dynamics on the weighted projected network, and the corresponding PW voter dynamics on the projected network. While the equivalence between the HO and PW dynamics follows directly from the mapping itself, the comparison with the corresponding PW voter dynamics allows us to assess the role played by the weights induced by the HO interactions and the question of macroscopic reducibility.

This comparison is performed at the macroscopic level, focusing on two observables that characterize the ordering process:
\begin{itemize}
 \item \emph{Density of active links} $\rho$: In complex networks, this quantity is defined as the fraction of links connecting nodes in opposite states. For hypergraphs, $\rho$ corresponds to the density of active links in the associated projected network.
 \item \emph{Fixation time} $\tau$: The time required to reach one of the two absorbing consensus states ($x=0$ or $x=1$), starting from a given initial configuration.
\end{itemize}

In addition, for the hypergraph-linear voter model we analytically demonstrate—via a pair approximation—that the ordering dynamics does not depend on the specific heterogeneity of the weights defining the PW dynamics. In other words, at the macroscopic level, the hypergraph-linear voter model is dynamically equivalent to the voter model on the projected network.

\begin{figure*}
 \centering
 \includegraphics[width=0.49\linewidth]{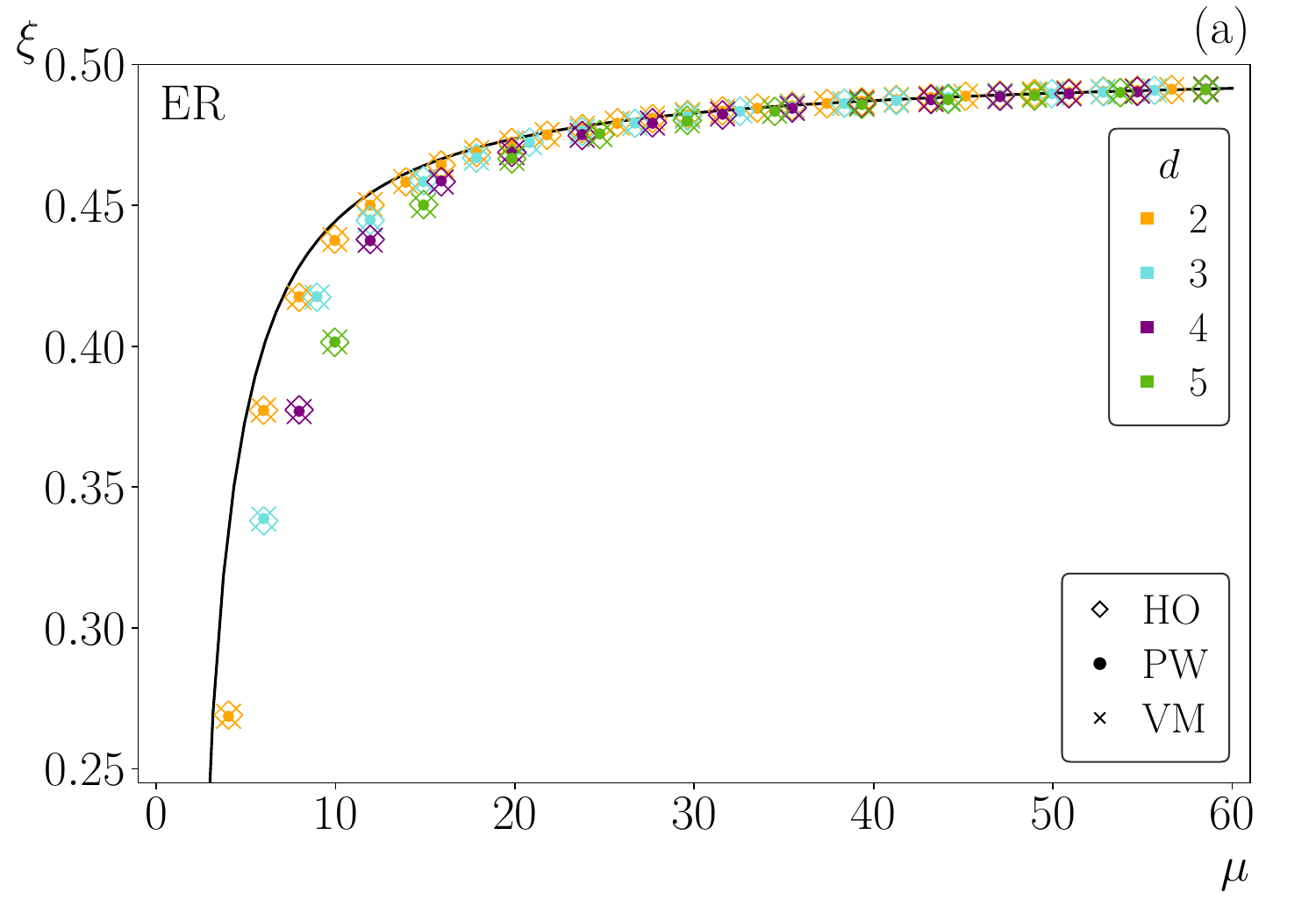}
 \includegraphics[width=0.49\linewidth]{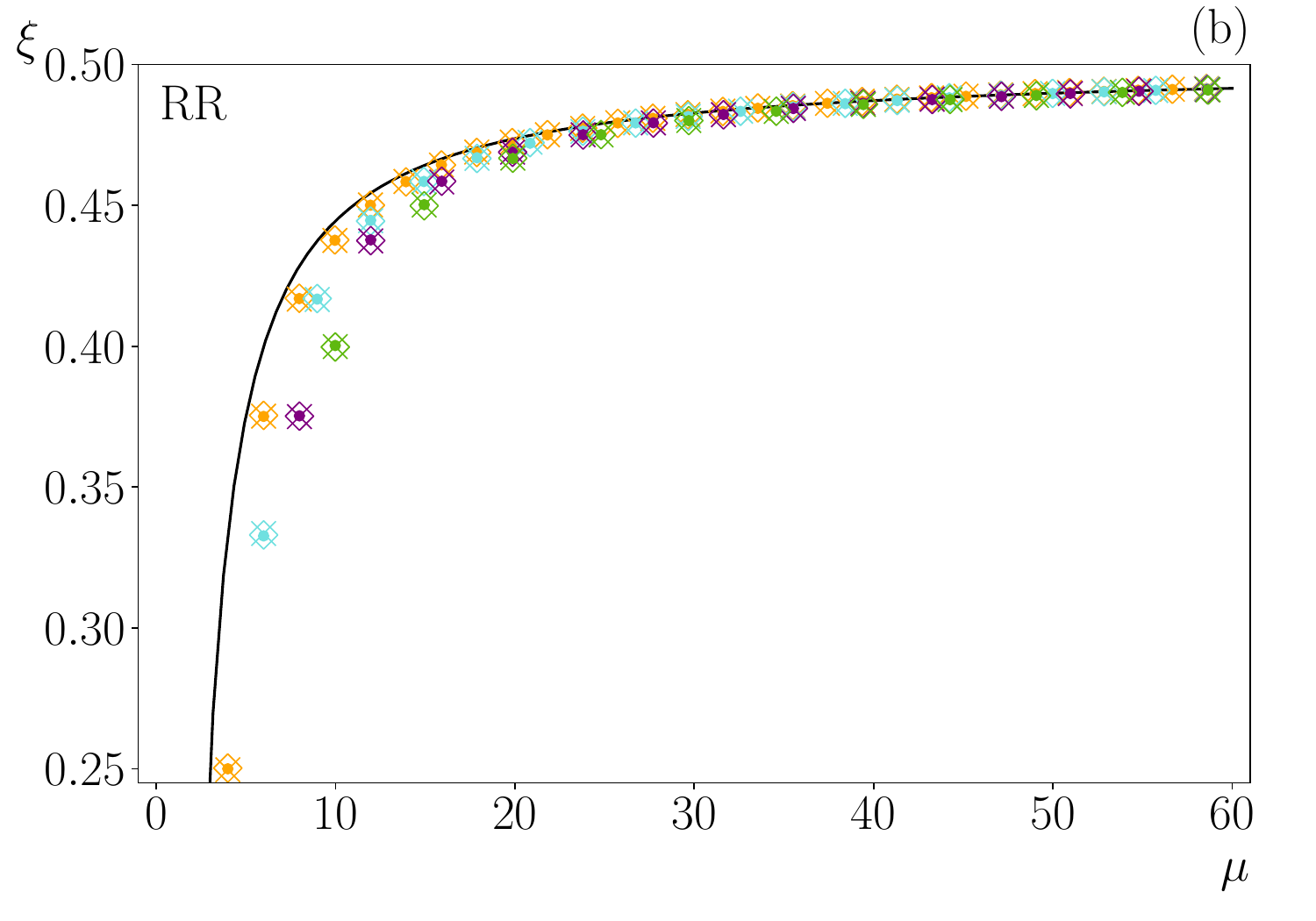}
 \includegraphics[width=0.49\linewidth]{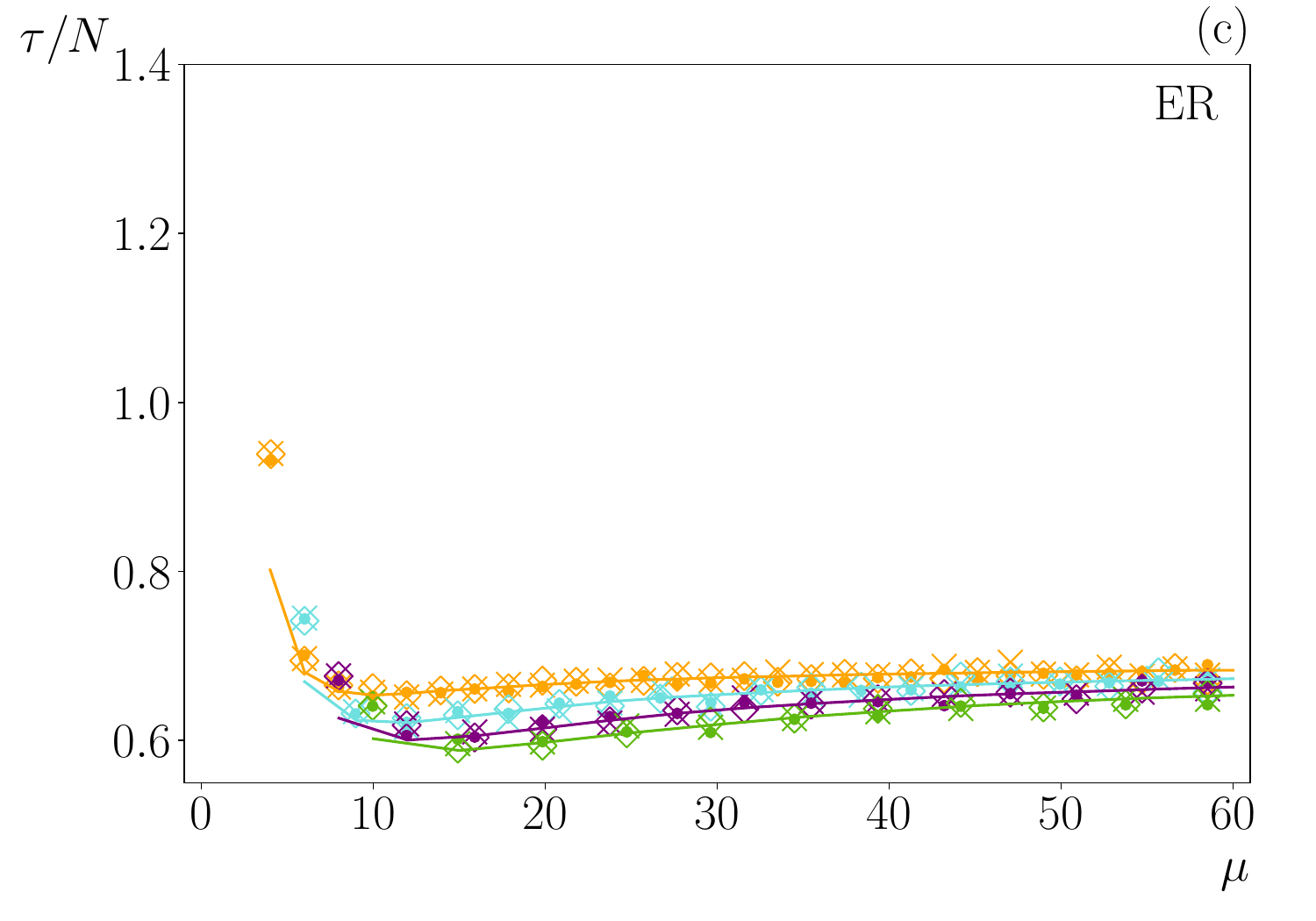}
 \includegraphics[width=0.49\linewidth]{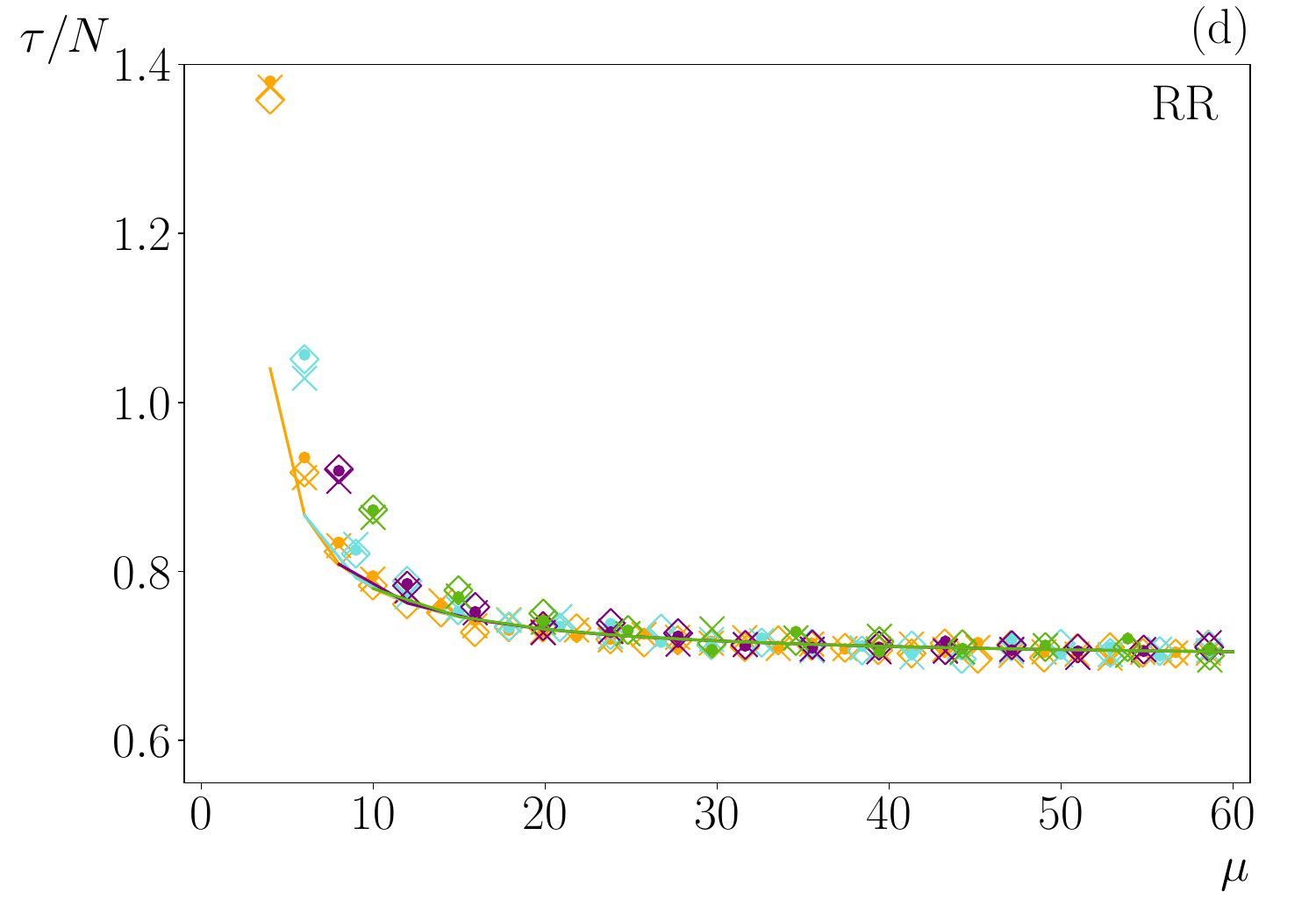}
 \caption{\textbf{Hypergraph-linear voter model.} (a,b) Plateau value $\xi$, determined as the temporal average of the ratio $\rho(t)/[4x(t)(1-x(t))]$, and (c,d) scaled fixation time $\tau/N$ versus the average degree of the projected network $\mu$, for Erd\H{o}s--R\'enyi (ER) and $z$-regular random (RR) $d$-hypergraphs, with initial condition $x(0)=1/2$. We compare three dynamics: higher–order on hypergraphs (HO), reduced pairwise dynamics on the weighted projected network (PW), and voter model on the projected network (VM). The three dynamics overlap. Solid lines correspond to theoretical predictions from the pair approximation, Eqs.~\eqref{eq:plateau_PA} and \eqref{eq:tau_PA}, respectively. Legends are shown once for clarity. The system size is $N=12000$ for (a, b), and $N=1200$ for (c,d).}
 \label{fig:VM}
\end{figure*}

\subsubsection{Hypergraph-linear voter model} \label{sec:Map_VM}

The hypergraph-linear voter model corresponds to the linear case $q=1$ of the social impact function defined in Eq.~\eqref{eq:SIF_GVM}.

For the fully connected hypergraph and the complete $d$-hypergraph, substituting the linear form into the respective expressions for the weights, Eqs.~\eqref{eq:weights-FCH} and \eqref{eq:w_CH}, yields the same state-independent weights,
\begin{equation} \label{eq:w_VM}
 \omega(s_i,s_j) = \frac{1}{N-1},
\end{equation}
for all $s_i,s_j=\pm1$. These weights correspond to the standard voter model dynamics on the complete graph. In other words, on these homogeneous hypergraphs the hypergraph-linear voter model maps exactly onto the voter model on the complete graph~\cite{VM_CG}.

In the case of arbitrary hypergraphs, substituting the social impact function in Eq.~\eqref{eq:SIF_GVM} with $q=1$ into Eqs.~\eqref{eq:wij_ei_flip} and \eqref{eq:wij_ei} yields the following weights for each hyperedge $e$
\begin{eqnarray} \label{eq:wij_ei_VM}
 \omega_i^e(s_i,s_j)= \frac{1}{\kappa_i d_e },
\end{eqnarray}
which lead to the total weight
\begin{eqnarray} \label{eq:wij_VM}
 \oij = \frac{1}{\kappa_i} \sum_{e \in \mathcal{E}_i \mid j \in e} \frac{1}{d_e}. 
\end{eqnarray}
We note that even in arbitrary hypergraphs, state-independent weights are obtained. These weights can be interpreted as follows: the probability of interaction of a node $i$ is distributed across all its neighbors by a two-stage uniform split: 
\begin{enumerate}
 \item Node $i$ divides a unit weight equally among its $\kappa_i$ incident hyperedges, so each incident hyperedge $e$ receives $1/\kappa_i$.
 \item Within each such hyperedge $e$, this amount is distributed equally among the $d_e$ neighbors of $i$ in $e$. Therefore, each neighbor $j$ in $e$ gets a weight $\omega_{ij}^e=(\kappa_id_e)^{-1}$.
\end{enumerate}
If $i$ and $j$ share multiple hyperedges, the contributions add up. 

Unlike the fully connected and complete hypergraph cases, where both descriptions reduce to the VM on the complete graph, the presence of structural heterogeneities yields non-uniform pairwise interactions. Although closed-form expressions for $\oij$ are generally unavailable in generic hypergraphs, explicit expressions can be derived in the particular case of RR $d$-hypergraphs, see Appendix~\ref{sec:app:weights}.

We show now—via a pair approximation—that, in the hypergraph-linear voter model, the heterogeneity of the weights does not affect the macroscopic dynamics.

\paragraph*{\textbf{Pair approximation}\\}

An important advantage of the mapping of HO dynamics into a PW dynamics is that it enables analytical treatment through well-established theoretical approaches. In particular, the voter model on complex networks admits accurate approximations that capture dynamical correlations with different levels of detail, such as the pair approximation~\cite{Vazquez_PA}, the heterogeneous pair approximation~\cite{Pugliese_2009}, and approximate master equations~\cite{Gleeson}. These frameworks allow one to derive explicit expressions for macroscopic observables.

As developed, these theories assume that all neighbors contribute equally to the dynamics, corresponding to degree-homogeneous weights ($\oij=1/k_i$ for $j \in \mathcal{V}_i$). In contrast, the PW dynamics defined in Sec.~\ref{sec:general_pw} displays a heterogeneous weight distribution. To properly account for this heterogeneity, we develop in Appendix~\ref{app:PA} a pair approximation for uncorrelated weighted networks (PA), extending the classical formulation to the case of an arbitrary weight distribution. Based on the result obtained for the hypergraph-linear voter model in Eq.~\eqref{eq:wij_VM}, we assume that the weights are state-independent.

Remarkably, the resulting evolution equations for $x$ and $\rho$ turn out to be identical to those of the degree-homogeneous case~\cite{Vazquez_PA}, namely
\begin{subequations} \label{eq:rate_eq_PA}
 \begin{align}
 \frac{dx}{dt} &= 0,\\[4pt]
 \frac{d\rho}{dt} &= \frac{2\rho}{\mu}
 \left[ (\mu-1)\!\left(1-\frac{\rho}{2x(1-x)}\right) - 1 \right],
 \label{eq:drhodt_VM_PA}
 \end{align}
\end{subequations}
where we recall that $\mu$ is the average degree of the projected network. Hence, the PW dynamics on an uncorrelated, state-independent weighted network is equivalent to the corresponding dynamics on an uncorrelated degree-homogeneous weighted network.

This dynamical system presents two stationary solutions: the trivial solution $\rho^*=0$, corresponding to the consensus states, which is stable for $\mu<2$, and a nontrivial solution that is stable for $\mu>2$. Since $x$ is conserved, this nontrivial solution is fully determined by the initial condition $x_0 \equiv x(0)$, namely
\begin{equation}
 \rho^*(x_0) = 4\xi(\mu)\,x_0(1-x_0),
 \label{eq:rho_PA}
\end{equation}
where we have defined
\begin{equation} \label{eq:plateau_PA}
 \xi(\mu) \equiv \frac{\mu-2}{2(\mu-1)}.
\end{equation}
Equations~\eqref{eq:rate_eq_PA} describe the deterministic dynamics in the thermodynamic limit ($N\to\infty$), where stochastic fluctuations vanish and $x$ is conserved. In this limit, the system rapidly relaxes onto the manifold defined by Eq.~\eqref{eq:rho_PA}.

In finite systems, however, the trajectories of the system in phase space $(x,\rho)$ fluctuate around the parabola $\rho(x)=4\xi(\mu)x(1-x)$  until it eventually reaches one of the absorbing states, $x=0$ or $x=1$~\cite{Vazquez_PA}. The parameter $\xi(\mu)$ determines the maximum of this parabola at $x=1/2$, setting the characteristic plateau value of the density of active links.

Given that the heterogeneity of the weights does not affect the macroscopic dynamics within the pair-approximation description, we expect that the results derived in Ref.~\cite{Vazquez_PA} for finite systems remain valid. In particular, the fixation time starting from an initial fraction $x_0$ is given by
\begin{equation} \label{eq:tau_PA}
 \tau = -\,\frac{(\mu-1)\mu^2}{(\mu-2)\mu_2}\,
 N\,\big[\,x_0 \ln x_0 + (1-x_0)\ln(1-x_0)\,\big],
\end{equation}
where $\mu_2$ is the second moment of the degree distribution $P(k)$ of the network.

As an illustration of the derived results, we perform numerical simulations of the HO dynamics on ER and RR $d$-hypergraphs for several values of the hyperedge order $d$ and the average hyperdegree $\mu_d$. We compare these results with those obtained from the PW dynamics on the corresponding weighted projected networks, with weights given by Eq.~\eqref{eq:wij_VM}, as well as with the standard voter model on the projected networks (VM), corresponding to degree-homogeneous weights.

We focus on two main observables. First, we compute the plateau value $\xi$, defined as the temporal average of the ratio $\rho(t)/[4x(t)(1-x(t))]$, evaluated along dynamical trajectories that have not yet reached an absorbing state. Second, we measure the fixation time $\tau$, considering a symmetric initial condition with $x_0=1/2$. In Fig.~\ref{fig:VM}, we show the plateau $\xi$ and the fixation time $\tau$ as a function of the average degree of the projected network $\mu$, which depends on both the hyperedge order $d$ and the hypergraph average degree $\mu_d$. Whereas the perfect agreement between the HO and PW dynamics is expected from the exact mapping developed in Sec.~\ref{sec:reduc-HG}, what is remarkable is that these two dynamics also coincide with that of the VM on the projected network, as predicted by the PA.

Additionally, we compare the numerical simulations with the analytical results of the PA, Eqs.~\eqref{eq:plateau_PA} and~\eqref{eq:tau_PA}. We see that for moderate and high values of the mean degree $\mu$, the numerical results are well approximated by the theoretical prediction given by the PA. However, for small $\mu$ the measured plateau values lie systematically below the PA curve. This discrepancy originates from structural correlations induced by the hypergraph construction: when hyperedges are projected into links, closed motifs such as triangles or squares are formed by definition, generating clustering and dynamical correlations between second nearest neighbors that are neglected by the PA. As shown in Ref.~\cite{Gleeson_corr}, such correlations reduce the value of the plateau, precisely as observed in our simulations. For the same reason, consensus times are systematically larger than the theoretical estimates at low $\mu$. Therefore, the deviations from the PA at small $\mu$ highlight its limitations in the presence of clustering.

In summary, these results demonstrate that the macroscopic description of the hypergraph-linear voter model dynamics is effectively insensitive to higher-order topologies, as predicted by the PA. The explicit form of the weights in the projected network resulting from the mapping becomes irrelevant, and all relevant structural information is encoded in the degree distribution of the projected network.

\begin{figure*}[t]
 \centering
 \includegraphics[width=0.49\textwidth]{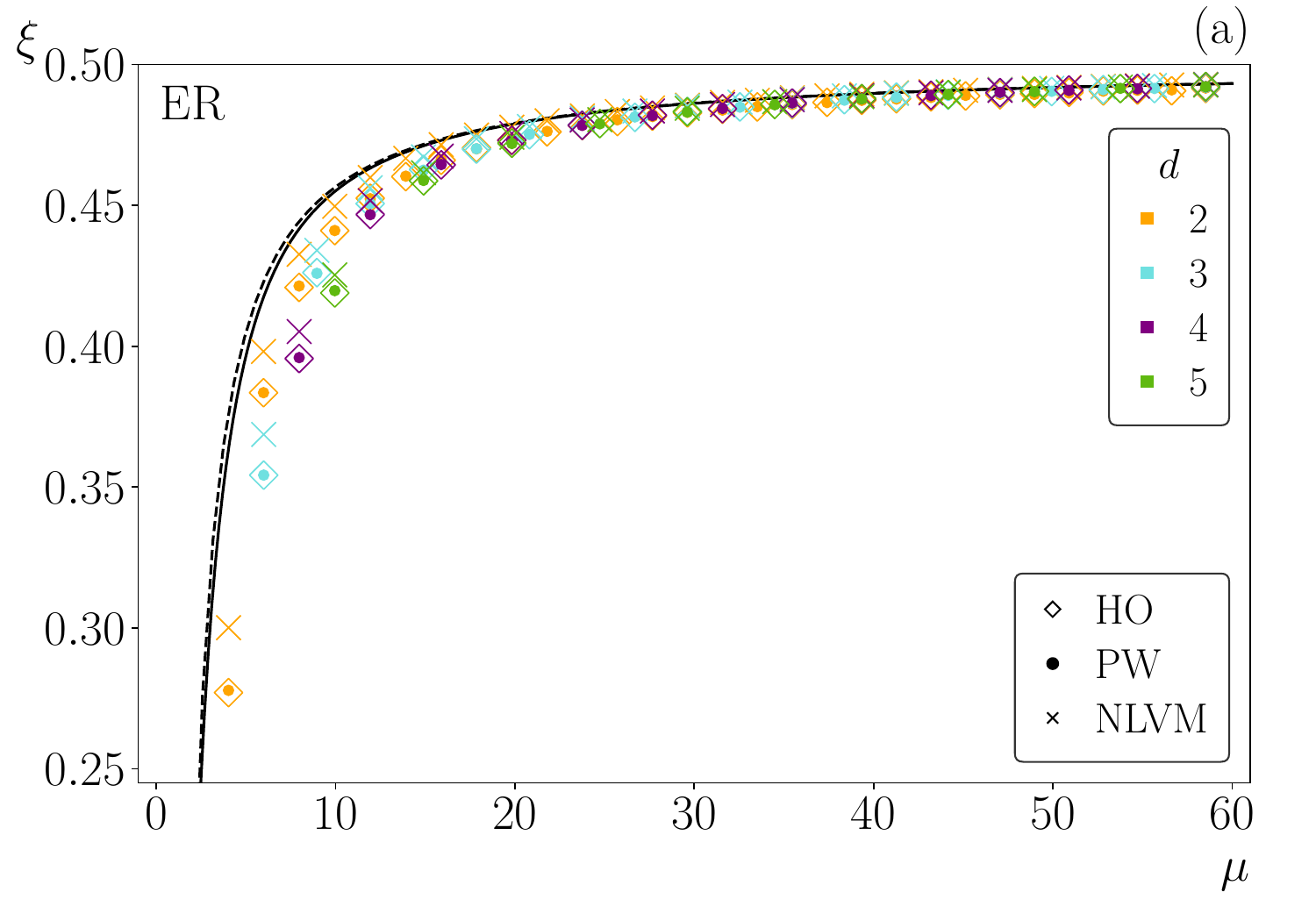}
 \includegraphics[width=0.49\textwidth]{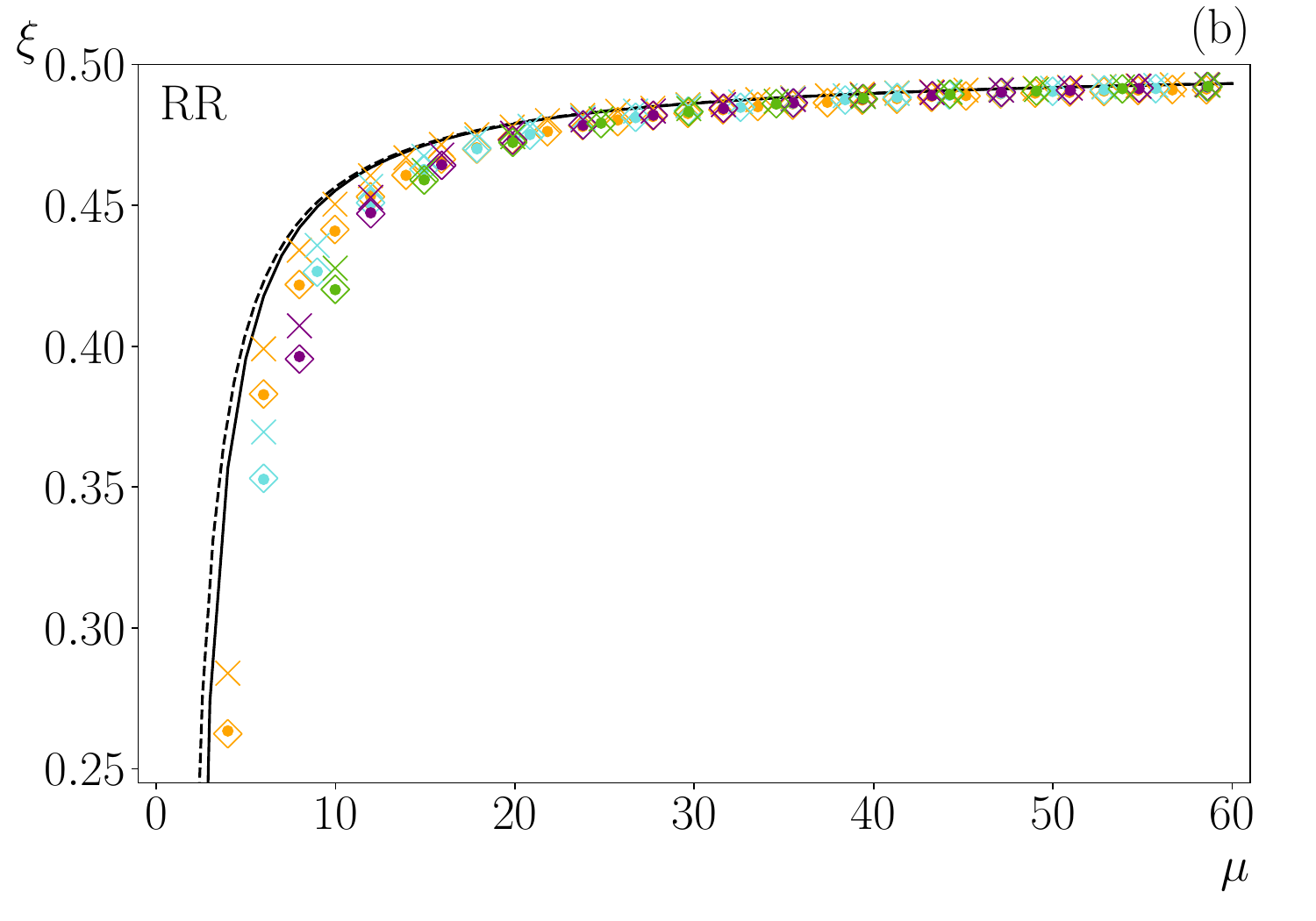}
 \caption{\textbf{Hypergraph-nonlinear voter model ($\boldsymbol{q=0.8}$).} Plateau value $\xi$ versus the average degree of the projected network $\mu$, for (a) Erd\H{o}s--R\'enyi (ER) and (b) $z$-regular random (RR) $d$-hypergraphs. We compare three dynamics: higher–order on hypergraphs (HO), reduced pairwise dynamics on the weighted projected network (PW), and nonlinear voter model on the projected network (NLVM). Results for HO and PW overlap. Solid lines represent theoretical predictions from the pair approximation for the NLVM~\cite{Ramirez2024} while dashed lines correspond to the approximate expression $\rho^*=(\mu-2q)/[2(\mu-q)]$~\cite{Min2017, Jedr_PA}. Legends are shown once for clarity. Parameter values $N=12000$, $q=0.8$.}
 \label{fig:NLVM_q<1}
\end{figure*}

\subsubsection{Hypergraph-nonlinear voter model}\label{sec:map_CH_NLVM}

For the fully connected hypergraph, the weights of the projected network obtained by substituting Eq.~\eqref{eq:SIF_GVM} into Eq.~\eqref{eq:weights-FCH} read
\begin{subequations}
 \begin{align}
 \omega(-,+) &\simeq \frac{x^{q-1}}{N}, \\
 \omega(-,-) &\simeq \frac{1-x^q}{N(1-x)}, \\
 \omega(+,-) &\simeq \frac{(1-x)^{q-1}}{N}, \\
 \omega(+,+) &\simeq \frac{1-(1-x)^q}{Nx}.
 \end{align} 
 \label{eq:weights-FCH-2}
\end{subequations}

For the complete $d$-hypergraph, the weights of the projected network obtained by substituting Eq.~\eqref{eq:SIF_GVM} into Eq.~\eqref{eq:w_CH} are given by
\begin{subequations}
 \begin{align}
 \omega(-,+)&= \frac{\langle n^q \rangle}{Nxd^q}, \\
 \omega(-,-)&= \frac{d^q-\langle n^q \rangle}{N(1-x)d^q}, \\
 \omega(+,-)&= \frac{\langle (d-n)^q \rangle}{N(1-x)d^q}, \\
 \omega(+,+)&= \frac{d^q-\langle (d-n)^q \rangle}{Nxd^q},
 \end{align}
 \label{eq:w_CH_NLVM}
\end{subequations}
where the average $\langle \cdot \rangle$ is defined as in Eq.~\eqref{eq:Fdx}. Note that these weights coincide with those of the fully connected hypergraph, Eq.~\eqref{eq:weights-FCH-2}, in the limit $d=N-1$.

\begin{figure*}[t]
 \centering 
 \includegraphics[width=0.49\linewidth]{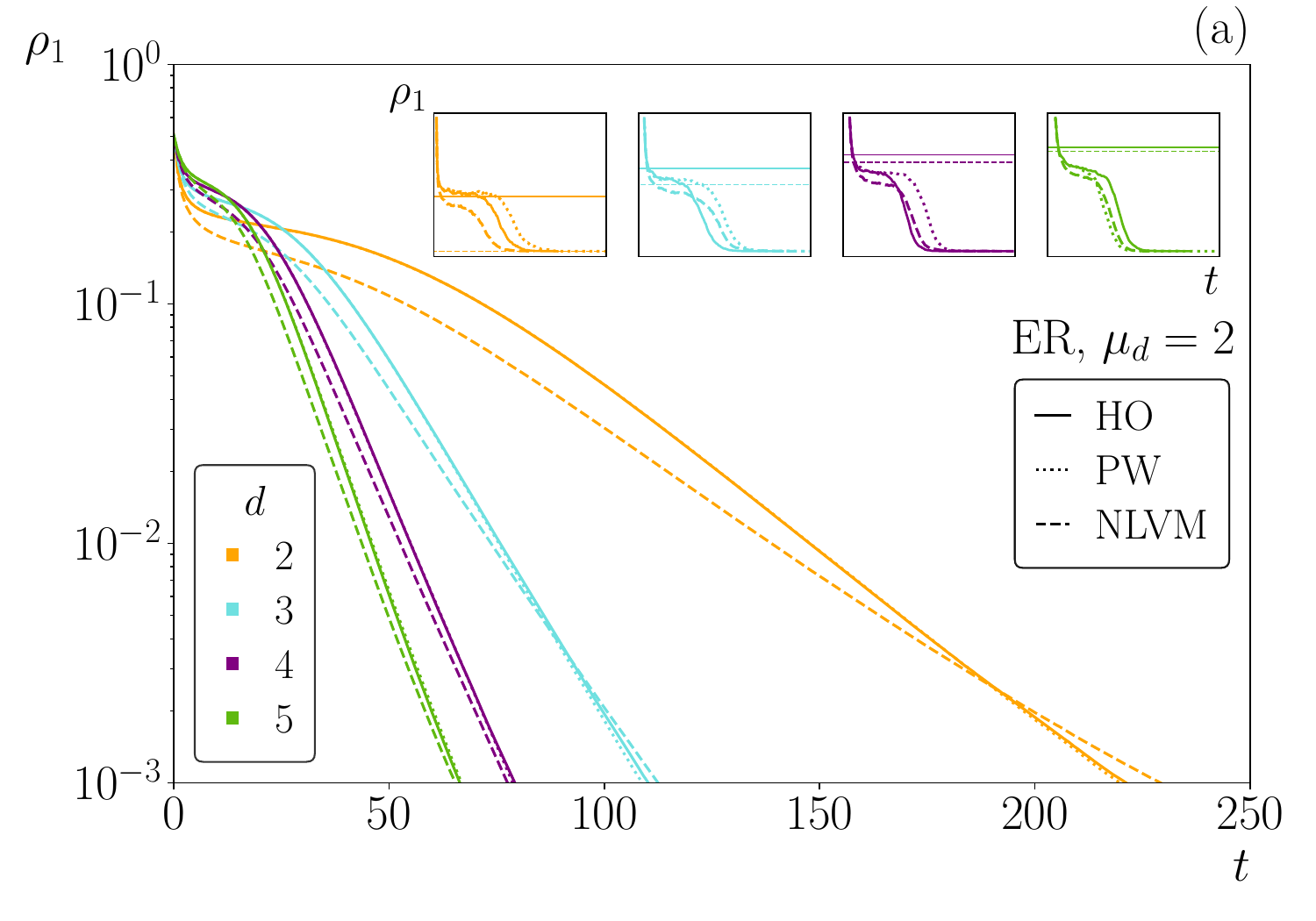}
 \includegraphics[width=0.49\linewidth]{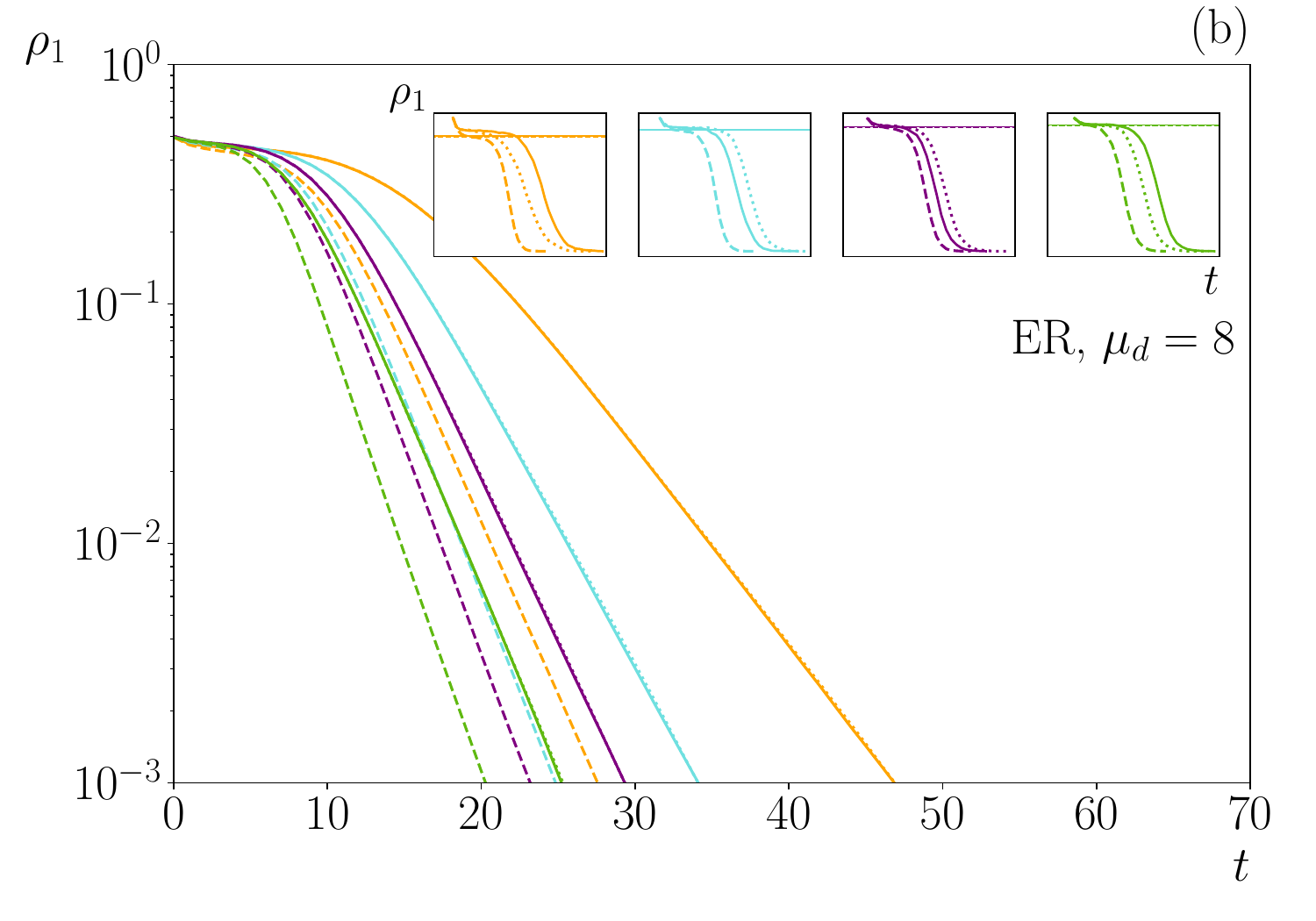}
 \includegraphics[width=0.49\linewidth]{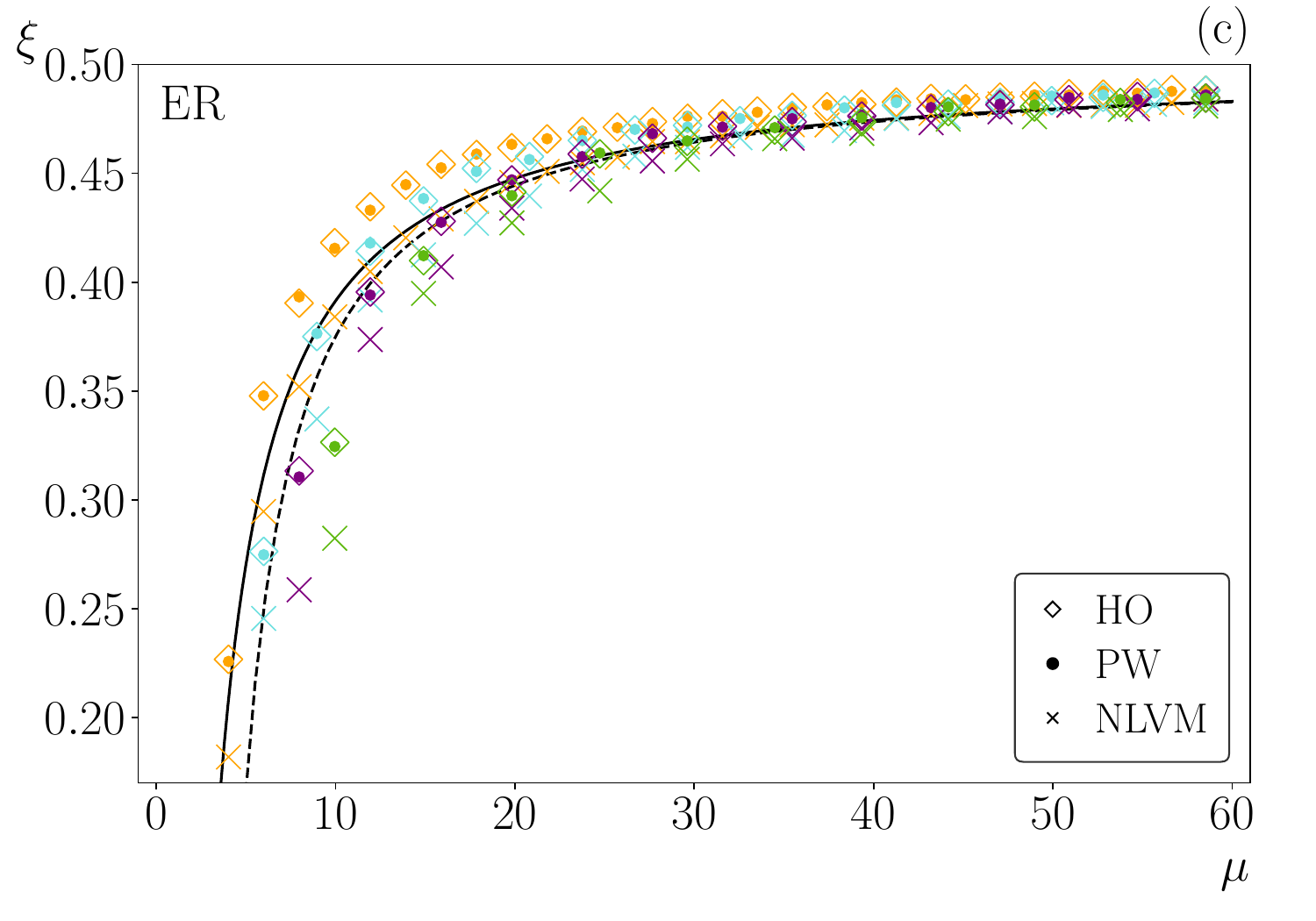}
 \includegraphics[width=0.49\linewidth]{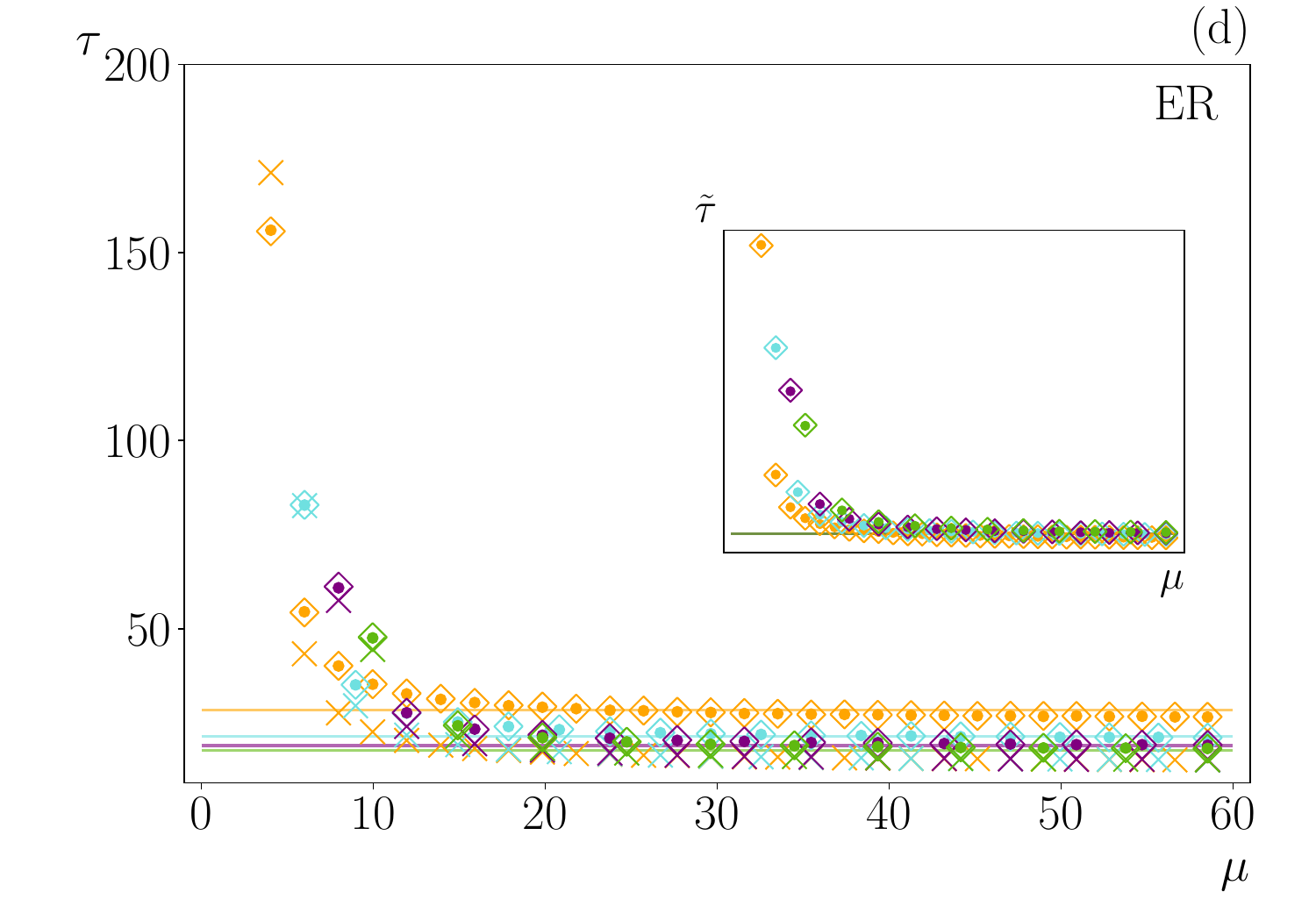}
 
 \caption{\textbf{Hypergraph-nonlinear voter model ($\boldsymbol{q=2}$).} (a, b) Time evolution of the average density of active links $\rho$ for Erd\H{o}s--R\'enyi (ER) $d$-hypergraphs with average hyperdegree $\mu_d=2,8$, respectively. Insets: Single trajectories for each $d$. (c) Plateau value $\xi$ versus the average degree of the projected network $\mu$ for ER $d$-hypergraphs. Solid line correspond to the plateau values given by the pair approximation developed in Ref.~\cite{Ramirez2024} while dashed line correspond to the approximate expression $\rho^*=(\mu-2q)/2(\mu-q)$~\cite{Min2017, Jedr_PA}. (d) Fixation time versus the average degree of the projected network $\mu$ for ER $d$-hypergraphs. Horizontal solid lines correspond to the fixation times for the complete $d$-hypergraph obtained in the limit $N\to\infty$ from Ref.~\cite{Min2025}, $\tau\sim 2 d \ln N /(d-1)$. Inset: Rescaled fixation time $\tilde{\tau}\equiv (1 - 1 / d)\, \tau$ versus the average degree of the projected network $\mu$ for HO and PW dynamics. We compare three dynamics: higher–order interactions on hypergraphs (HO), reduced pairwise dynamics on the weighted projected network (PW), and nonlinear voter model on the projected network (NLVM). Results for HO and PW overlap. Legends are shown once for clarity. Parameter values: $N=1200$ for (a, b, d) and $N=12000$ for (c).}
 \label{fig:NLVM_q=2}
\end{figure*}

For arbitrary hypergraphs, the substitution of Eq.~\eqref{eq:SIF_GVM} into Eq.~\eqref{eq:wij_ei_flip} yields the following weights for nodes in the opposite state in each hyperedge $e$
\begin{equation} \label{eq:wij_ei_NLVM}
 \omega^e_i(s_i,-s_i) =
 \begin{cases}
 \dfrac{1}{\kappa_i n_e}\left(\dfrac{n_e}{d_e}\right)^q, & \text{if } s_i=-1,\\[4pt]
 \dfrac{1}{\kappa_i (d_e - n_e)}\left(1-\dfrac{n_e}{d_e}\right)^q, & \text{if } s_i=+1.
 \end{cases}
\end{equation}
In contrast to the linear case, the resulting weights depend explicitly on the local state configuration and therefore evolve in time. The total weight $\oij$ is obtained with Eq.~\eqref{eq:wij} by summing over all the contributions $\omega_i^e(s_i,s_j)$.

In our numerical simulations, we observe that the dynamics of the HO model on random $d$-hypergraphs, such as ER and RR $d$-hypergraphs, is qualitatively similar to that observed for the complete $d$-hypergraph discussed in Sec.~\ref{sec:HGVM_dynamics}. For $q<1$, the system reaches a stable coexistence state characterized by a plateau value $\rho^*=\xi$ and $x^*=1/2$, from which it eventually reaches one of the two absorbing states due to finite-size fluctuations. For $q>1$, the dynamics is instead characterized by two stable absorbing states. After a short transient, the system approaches one of the two consensus states, with the density of active links $\rho(t)$ decaying exponentially to zero. Because of this phenomenology we will discuss separately the cases $q<1$ and $q>1$. In addition, and motivated by the linear case, where the heterogeneity of the weight distribution was found to be irrelevant, we now compare the HO dynamics of the hypergraph-nonlinear voter model with its reduced PW dynamics on the projected network and with the NLVM, together with the theoretical predictions of the pair approximation for this latter case~\cite{Ramirez2024}.

In Fig.~\ref{fig:NLVM_q<1}, we show the plateau values $\xi$ obtained for $q=0.8$ on ER and RR $d$-hypergraphs. As predicted by the mapping, the plateau values resulting from the HO and PW dynamics overlap over the entire range of parameters. The NLVM provides an overall good description of the dependence of $\xi$ on the average degree $\mu$, becoming quantitatively accurate for sufficiently connected hypergraphs. In this regime, all three models yield a plateau value that is independent of the hyperedge order $d$ and coincides with the value predicted by the pair approximation for the NLVM~\cite{Ramirez2024}.

In Fig.~\ref{fig:NLVM_q=2} we present the results for $q=2$. Figures~\ref{fig:NLVM_q=2}(a,b) show the temporal evolution of the average density of active links $\rho(t)$ for different values of $d$ and $\mu$, with single-realization trajectories displayed in the insets. Figure~\ref{fig:NLVM_q=2}(c) reports the plateau values $\xi$ as a function of the average degree of the projected network $\mu$. The same considerations discussed for the sublinear case ($q<1$) apply to the plateau values, with good quantitative agreement between the NLVM and the HO and PW dynamics as $\mu$ increases. However, the temporal evolution of $\rho(t)$ exhibits systematic deviations in the NLVM. These discrepancies are further highlighted in Fig.~\ref{fig:NLVM_q=2}(d), which shows the fixation time $\tau$ as a function of $\mu$. 

Beyond these discrepancies, it is worth noting that in the limit $\mu\to\infty$ the fixation time $\tau$ converges to the value predicted in Ref.~\cite{Min2025} for complete $d$-hypergraphs and $N\gg1$. In this regime, the rescaled fixation times $\tilde{\tau} \equiv (1 - 1/d)\,\tau$ for HO and PW dynamics collapse for all $d$ (see inset). This is in agreement with Eq.~\eqref{eq:Hd_q=2} which implies that, in the thermodynamic limit, the effect of the hypergraph topology given by $ H_d(x,q)$ is just a change of the time scale. Additionally, we find that the fixation times of the NLVM overlap for all values of $d$ in this limit. Moreover, $\tau$ exhibits a non-monotonic dependence on the nonlinearity parameter $q$ (see Appendix~\ref{app:NLVM_HG}). In particular, there exists an optimal value $q^*$ that minimizes the fixation time, in agreement with the results reported in Ref.~\cite{Min2025} for integer values of $q$. The precise location of $q^*$, however, depends sensitively on the topological properties of the underlying hypergraph.

In summary, and additionally to the exact microscopic mapping of HO to the PW interactions, the NLVM on the projected network provides a good overall description of the hypergraph-nonlinear voter model, yielding an excellent effective account of the plateau values for sufficiently large values of $\mu$. From a minimal-model perspective, this implies that, when analyzing empirical or simulated data subject to finite measurement uncertainty, the NLVM defined on the projected network constitutes an appropriate effective description.

\section{Discussion and Conclusions}\label{sec:conclusions}

We have studied social impact models with higher-order (HO) interactions defined on general hypergraphs, focusing on the fundamental question of whether such dynamics can be reduced to effective pairwise (PW) interactions. In these binary state models, a node flips its state with a probability determined by a function of the fraction of nodes within the same hyperedge that are in the opposite state. We have shown that, for node-update dynamics and arbitrary social impact functions, the resulting HO dynamics on a general hypergraph can be mapped exactly onto a PW dynamics on a weighted projected network. In the PW dynamics the change of state of a node $i$ only depends on the states of node $i$ and a previously chosen neighbor $j$. This mapping establishes a microscopic reducibility of HO interactions to PW interactions, where the weights $\omega_{ij}$ of the projected network fully encode the influence of higher-order structures by determining the probability that node $i$ interacts with node $j$. In general, the weights $\omega_{ij}$ depend implicitly on time through the evolving system configuration, making them difficult to compute explicitly.

For hypergraph voter models--defined by a power-law form of the social impact function--we explicitly computed the weights $\oij$ of the reduced PW dynamics for fully connected hypergraphs and complete $d$-hypergraphs, and we obtained them numerically for Erd\H{o}s--Rényi and $z$-regular random $d$-hypergraphs. For these models, we described the ordering dynamics in detail and, through numerical simulations, illustrated the equivalence between the original HO interactions dynamics on hypergraphs and the corresponding reduced PW dynamics on the weighted projected network.

Our results show that the nonlinearity of the social impact function plays a central role in determining the structure of the reduced PW dynamics. In the linear case ($q=1$), known as the \textit{hypergraph--linear voter model}, state changes correspond to random imitation of another node within the same hyperedge. This model therefore constitutes the natural hypergraph generalization of the classical voter model on complex networks~\cite{Castellano_2003,Suchecki_2004,Suchecki_2005,Sood_2005,Vazquez_PA}. In this regime, the reduced dynamics simplifies substantially: the weights $\oij$ for the projected network depend solely on topological properties of the hypergraph, such as node hyperdegrees and order of the hyperedges, and become independent of the system configuration. As a consequence, the weights are constant in time. 

This simplification enables the application of well-developed analytical methods for complex networks, which are largely unavailable for general hypergraphs. Using the PW reduced representation, we developed a pair approximation showing that the ordering dynamics is independent of the distribution of weights in the projected network and depends only on its degree distribution. Consequently, the macroscopic equations coincide with those of the standard voter model on the projected network with-degree homogeneous weights. While the pair approximation is quantitatively accurate only for sufficiently large mean degree $\mu$, its central prediction—namely, the independence of the macroscopic dynamics from weight heterogeneity—is confirmed numerically across the full parameter range explored. Therefore, at the level of macroscopic reducibility, the hypergraph--linear voter model on Erd\H{o}s--Rényi and $z$--regular random $d$--hypergraphs is equivalent to the standard voter model on the corresponding projected networks.

For $q\neq1$, the \emph{hypergraph-nonlinear voter model} generalizes to hypergraphs the nonlinear voter model (NLVM) defined on a complex network~\cite{Ramirez2024,Tobias2025,Llabres2025}, where the flipping probability depends on a power $q$ of the density of neighboring nodes in the network which are in the opposite state. In this nonlinear regime, the reduced PW dynamics unfolds on a temporal network whose link weights depend on the instantaneous configuration of node states. Beyond the question of exact microscopic reducibility, we also addressed the problem of identifying minimal models capable of reproducing macroscopic observables within a desired accuracy. We found that, for Erd\H{o}s--Rényi and $z$-regular random $d$-hypergraphs, the NLVM on the projected network gives an overall good description of the dynamics. In particular, the plateau value of the density of active links exhibited by the hypergraph-nonlinear voter model is well approximated by the NLVM on the pojected network, specially for sufficiently connected projected networks. In this regime, results are also largely insensitive to the hyperedge order $d$.

As a direction for future research, it would be valuable to examine social impact models on hypergraphs with hyperedge-update dynamics, where all nodes within a randomly selected hyperedge update their states simultaneously. By construction, these models cannot be mapped exactly onto a microscopically equivalent pairwise system, since pairwise models update only one node per interaction. However, macroscopic equivalence may still emerge. Indeed, this has been observed in the hypergraph-voter models on the complete $d$-hypergraph with hyperedge-update dynamics \cite{Min2025}. In the linear case ($q=1$), the macroscopic dynamics for $d>1$ is equivalent---up to a $d$--dependent time-rescaling factor---to that of the pairwise case ($d=1$), as described by a Fokker–Planck equation. This equivalence breaks down in the nonlinear case ($q \ne 1$), once again highlighting the central role of nonlinearity. More generally, identifying minimal or simplified models that are macroscopically equivalent to social impact models with hyperedge-update dynamics on general hypergraphs remains an open and compelling problem.

\begin{acknowledgments}
Partial financial support has been received from Grants PID2021-122256NB-C21/C22 and PID2024-157493NB-C21/C22 funded by MICIU/AEI/10.13039/501100011033 and by “ERDF/EU”, and the María de Maeztu Program for units of Excellence in R\&D, grant CEX2021-001164-M. We thank our colleagues Tobias Galla, Emilio Hernandez-Garcia, Lucas Lacasa and Sandro Meloni for useful discussions and comments.
\end{acknowledgments}

\bibliography{references.bib}
\clearpage

\appendix
\widetext
\renewcommand{\thefigure}{A\arabic{figure}}
\setcounter{figure}{0}


\section{Generation of Erd\H{o}s--R\'enyi $d$-hypergraphs}\label{sec:app:HG_generation}

We consider the generalization of the Erd\H{o}s--R\'enyi (ER) graphs~\cite{e1959,r1959, er1960} to $d$-hypergraphs~\cite{bookHG}. Given $N$ nodes and a target average hyperdegree $\mu_d$, we draw a total of $\mu_dN/(d+1)$ distinct $d$-hyperedges uniformly at random from the set of all possible $d$-hyperedges. This algorithm is used for most values of $\mu_d$. However, for small values of $\mu_d$ it does not guarantee that the resulting hypergraph is connected, i.e. not split into two or more disconnected sub-hypergraphs. A hypergraph is connected if its projected network is also connected.

A straightforward fix is simply to use the largest connected component. However, this alters both the size and the average hyperdegree. Instead, we here propose an alternative method for low values of $\mu_d$ for obtaining ER $d$-hypergraph with exactly $N$ nodes and average hyperdegree $\mu_d$ based on the relative size of the largest connected component $s(\mu_d)$. Assuming that $s(\mu_d)$ is the fraction of nodes which belong to at least one $d$-hyperedge, we can estimate the relative size of the largest connected component as
\begin{equation}
 s(\mu_d) \approx 1 - P_d(\kappa=0)=1 - \left( 1 - \frac{\mu_d}{\binom{N-1}{d}}\right)^{\binom{N-1}{d}} \approx 1 - e^{-\mu_d},
\end{equation}
where we have used the hyperdegree distribution of ER $d$-hypergraphs, Eq.~\eqref{eq:binomial}. 

Starting from an auxiliary hypergraph with average hyperdegree $\mu_0$, we want its largest connected component to have an average hyperdegree $\mu_d$. This condition leads to the relation
\begin{equation}
 \mu_d = \frac{\mu_0}{s(\mu_0)},
\end{equation}
whose explicit solution is given by 
\begin{equation}
\mu_0=\mu_d+W(-\mu_d e^{-\mu_d}).
\end{equation}
where $W(z)$ is the Lambert $W$ function~\cite{Lambert} and $\mu_0\ge 1$.

Given $\mu_0$, we then seek an auxiliary system size $N_0$ such that the
expected size of the largest connected component is the target $N$. This is done via a second empirical bisection: for each trial $N_0$ we generate several ER $d$-hypergraphs with parameters $(N_0,\mu_0)$, estimate the mean size of the largest connected component, adjusting $N_0$ until the largest connected component has size $N$.


\section{Degree distribution of the projected network of random $d$-hypergraphs}\label{sec:app:Mapping_HG}

\begin{figure*}[t]
 \centering
 \includegraphics[width=0.31\linewidth]{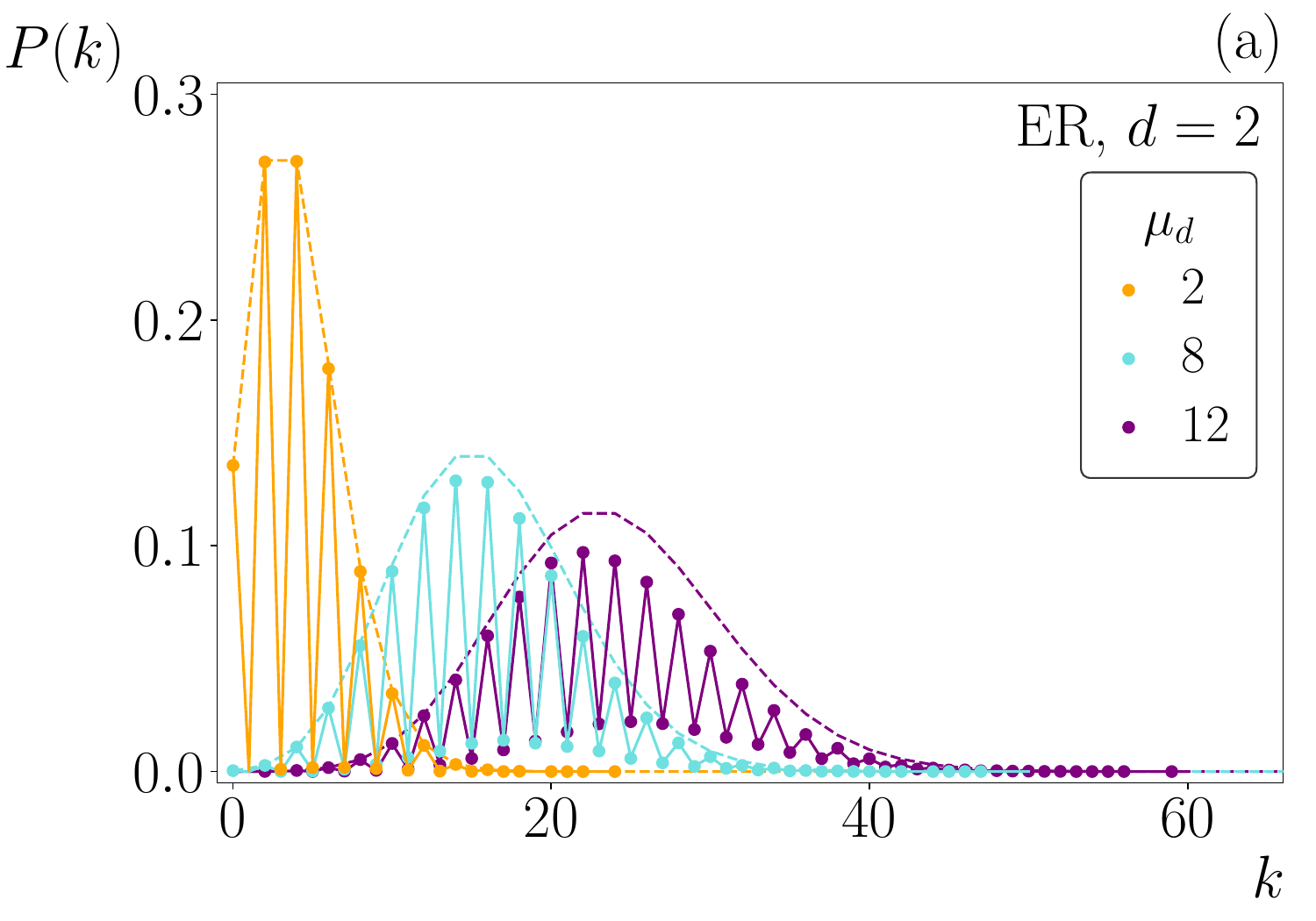}
 \includegraphics[width=0.31\linewidth]{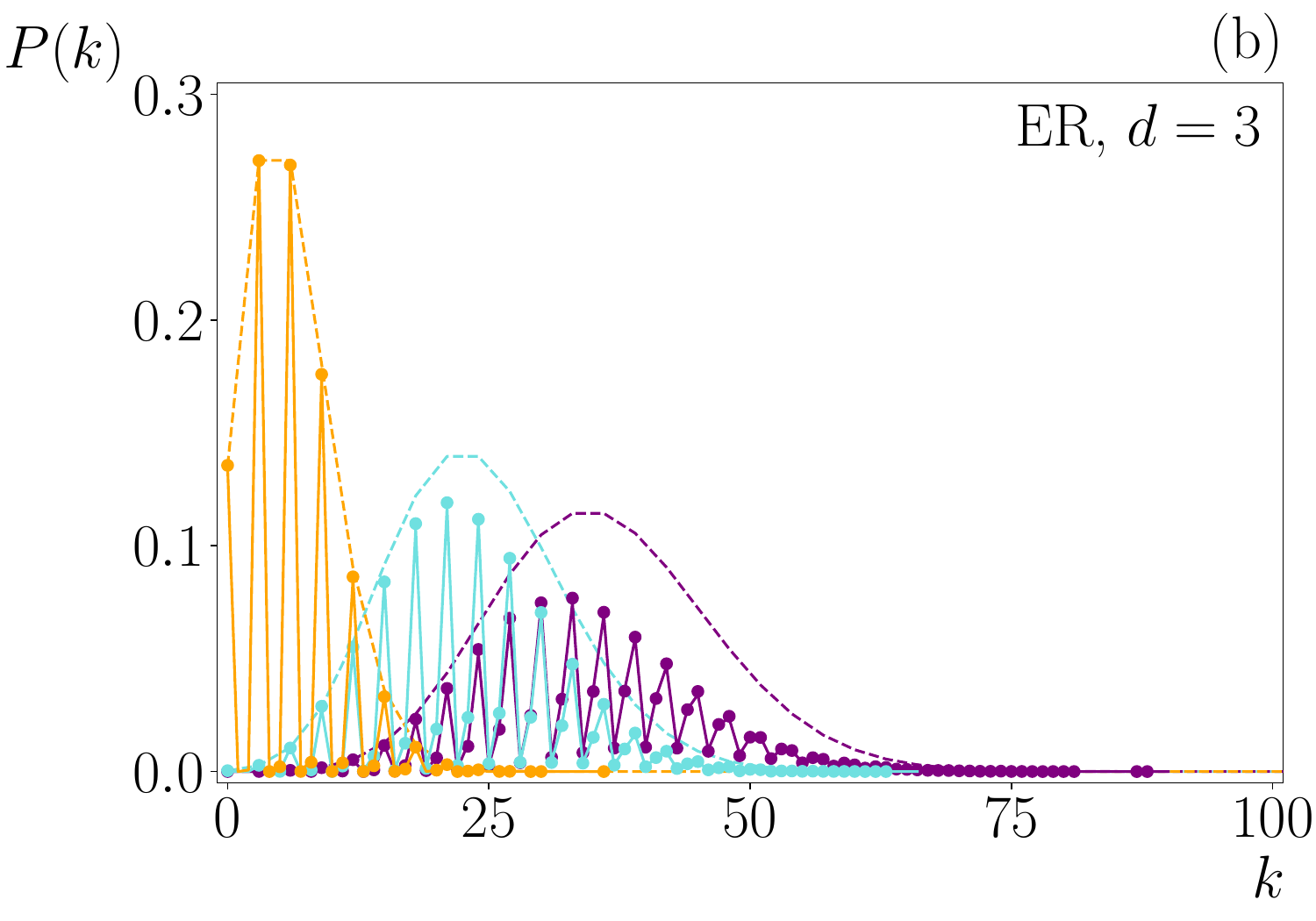}
 \includegraphics[width=0.31\linewidth]{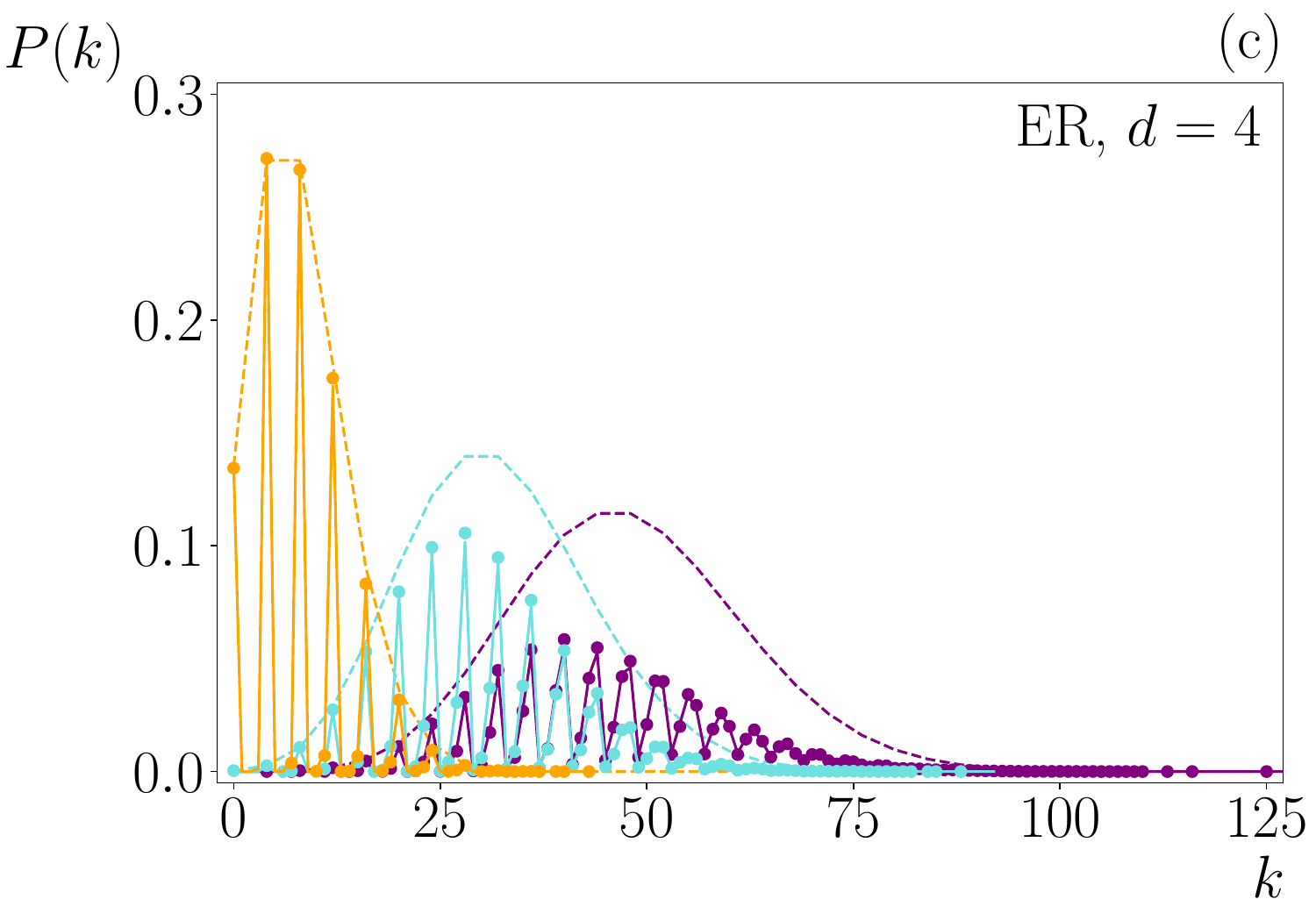}
 \includegraphics[width=0.31\linewidth]{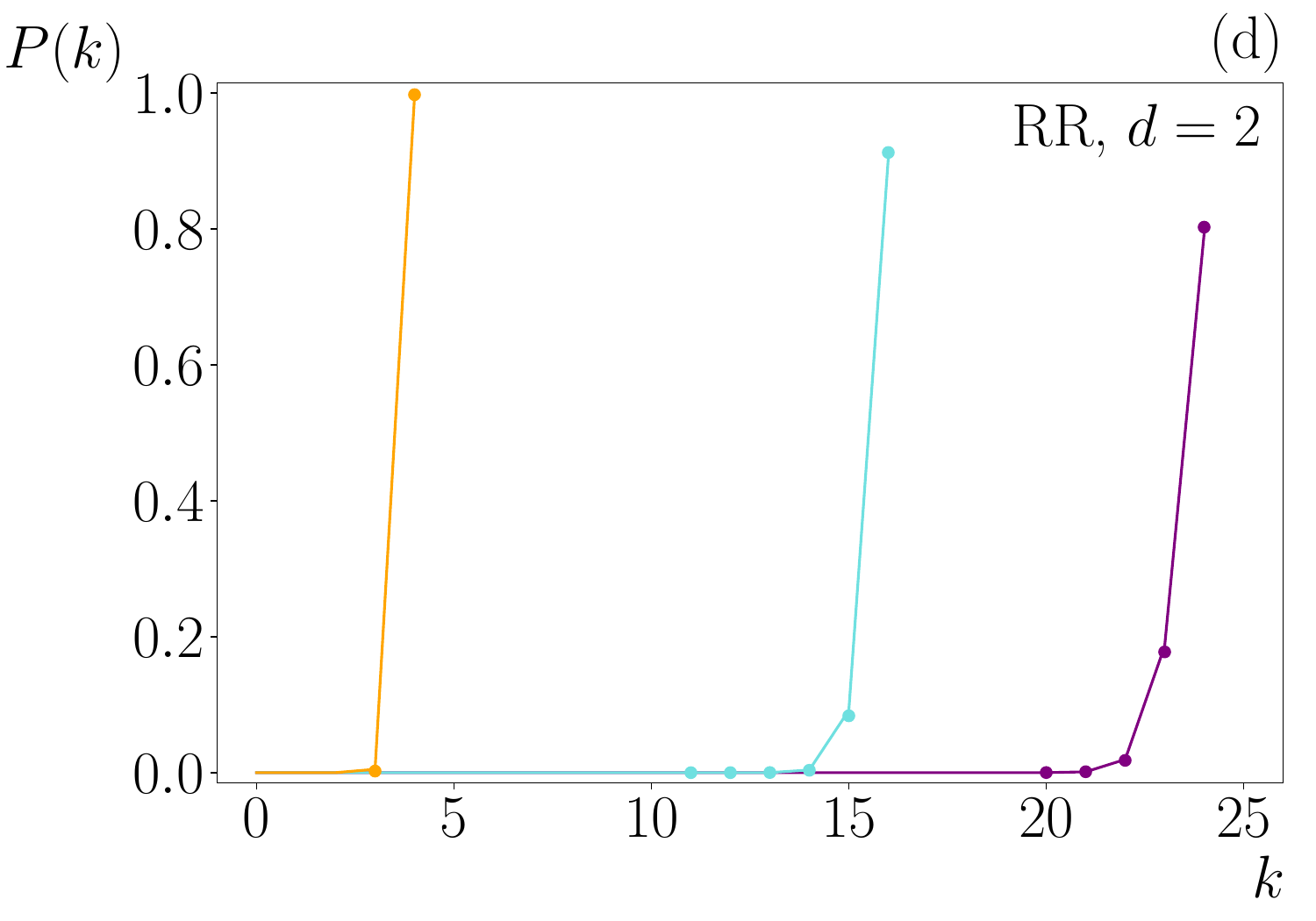}
 \includegraphics[width=0.31\linewidth]{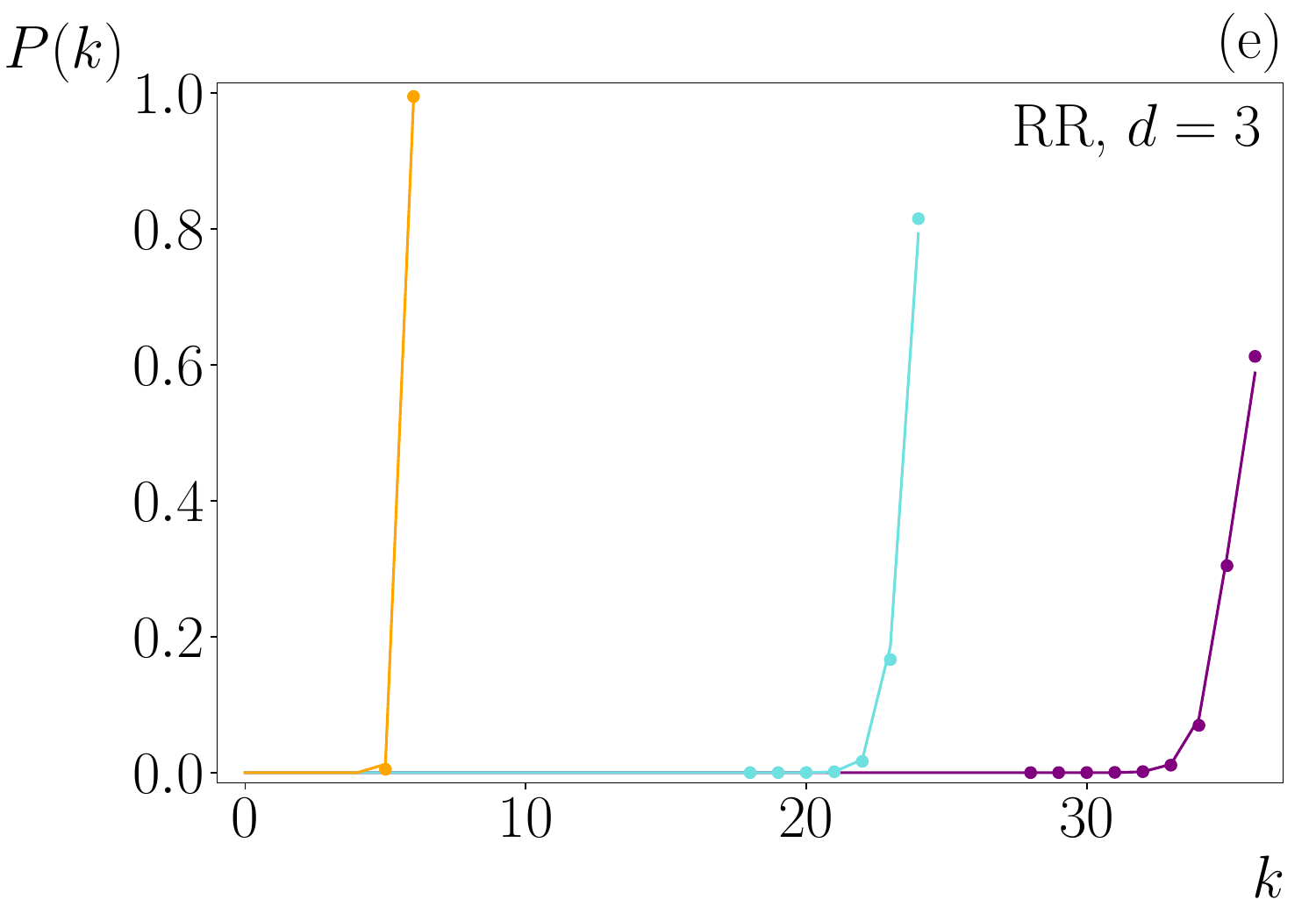}
 \includegraphics[width=0.31\linewidth]{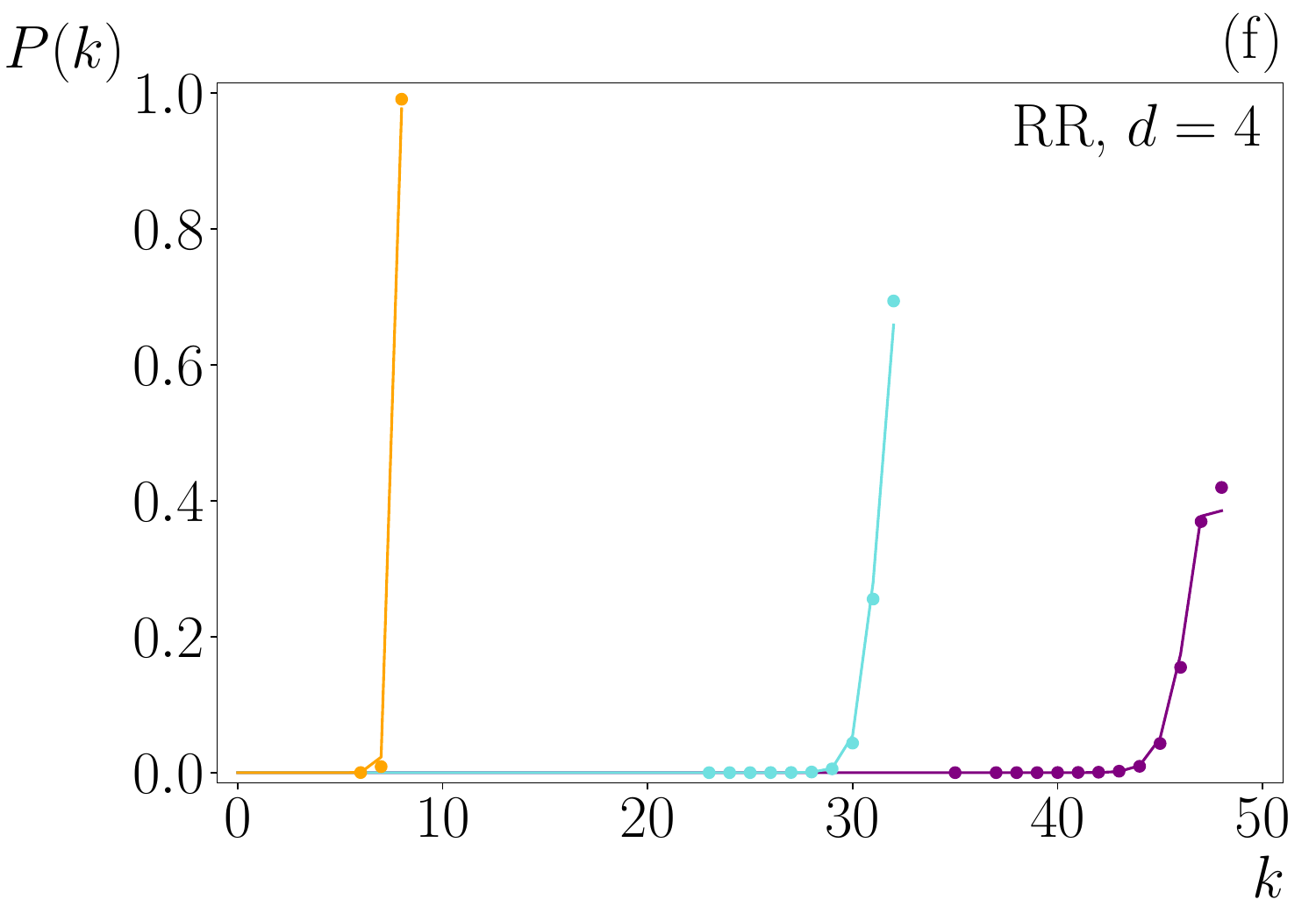}
 \caption{Degree distribution of the projected network of (a,b,c) Erd\H{o}s--R\'enyi (ER) and (d,e,f) $z$-regular random (RR) $d$-hypergraphs for several values of $d$. Symbols correspond to results obtained with computer simulations while solid lines correspond to the theoretical prediction Eq.~\eqref{eq:Pk_proj}. In (a,b,c), dashed envelope lines correspond to the theoretical curves in the sparse limit where the degree distribution is simply given by Eq.~\eqref{eq:binomial} with the rescaled horizontal axis $k=d\kappa$. Results of computer simulations are obtained averaging over $100$ $d$-hypergraphs of $N=1200$ nodes. Legend is shown once for clarity.}
 \label{fig:Pks}
\end{figure*}

We here determine the degree distribution $P(k)$ of the projected network, starting from the hyperdegree distribution $P_d(\kappa)$ of the underlying $d$-hypergraph.

In the limit of sparse hypergraphs, where it is improbable that two nodes share more than one hyperedge, repeated links in the projected network between any pair of nodes are very unlikely and each $d$-hyperedge contributes exactly $\binom{d+1}{2}$ distinct links between its constituent nodes. For a given node $i$, each hyperedge contributes to adding $d$ distinct neighbors. As a result, its degree is simply $ k_i = d\,\kappa_{i} $ and the degree distribution of the projected network remains invariant $P(k)=P_d(k/d)$. This simple relation holds only in the extreme sparse limit.

In general, overlap between hyperedges arises causing repetitions among neighbors and leading to $ k_i < d\,\kappa_{i} $. A related combinatorial approach was developed in Ref.~\cite{lopez_proj}, where an exact expression for the degree distribution of projected networks from Erd\H{o}s--R\'enyi $d$-hypergraphs is derived, explicitly accounting for overlaps between hyperedges. While exact, the method becomes computationally intensive for large systems. We therefore propose a simplified approximation that captures the essential effect of overlaps while offering a tractable alternative.

The degree distribution of the projected network is given by
\begin{equation} \label{eq:Pk_proj}
 P(k) = \sum_{\kappa} P_d(k \mid \kappa)\, P_d(\kappa),
\end{equation}
where $ P_d(k \mid \kappa) $ is the probability that a node participating in $ \kappa $ $d$-hyperedges, has exactly $ k $ unique neighbors in the projected network. The hyperdegree distribution $P_d(\kappa)$ depends on the particular topology of the $d$-hypergraph. For example, it is given by Eq.~\eqref{eq:binomial} for Erd\H{o}s--R\'enyi, while it reads $P_d(\kappa)=\delta_{\kappa, z}$ for $z$-regular random $d$-hypergraphs. 

Computing the exact probability $P_d(k \mid \kappa)$ is a non-trivial combinatorial problem since different hyperedges may share neighbors, producing correlated repetitions that depend on the overlap structure between hyperedges. To obtain an analytically tractable expression, we introduce the following approximation: sampling $\kappa$ hyperedges of size $d$, is equivalent to performing $d\kappa$ independent draws with replacement from the set of $N-1$ possible neighbors. We expect that this approximation is valid in the sparse limit. In this way, the quantity of interest becomes the probability that exactly $k$ distinct elements appear among these $d\kappa$ draws, which reads 
\begin{equation}
 P_d(k \mid \kappa) \;\approx\;
 \frac{
 \binom{N-1}{k}\; S(d\,\kappa,k) \; k! 
 }{
 (N-1)^{d\,\kappa}
 }.
 \label{eq:Pdkappa}
\end{equation}

This probability has a direct combinatorial interpretation. First, the factor $\binom{N-1}{k}$ counts the number of possible sets of $k$ neighbors. 
Second, the $d\kappa$ draws must be partitioned into $k$ non-empty unlabeled groups, each group representing the total draws that correspond to the same neighbor. The number of such partitions is given by the Stirling number of the second kind~\cite{AbramowitzStegun}
\begin{equation}\label{eq:stirling2}
 S(n,k)
 = \frac{1}{k!}
 \sum_{j=0}^{k}
 (-1)^{k-j}
 \binom{k}{j}
 j^{n}.
\end{equation}
The factor $k!$ accounts for the permutations arising from the fact that neighbors are distinguishable. Finally, the denominator $(N-1)^{d\kappa}$ is the total number of possible outcomes of the $d\kappa$ independent draws. 

Albeit approximate, the probability $P_d(k\mid \kappa)$ is properly normalized for each $\kappa$, satisfying $\sum_{k=1}^{d k_d} P_d(k \mid \kappa) = 1 $. In Fig.~\ref{fig:Pks}, we plot the degree distribution $P(k)$ of the projected network for Erd\H{o}s--R\'enyi and $z$-regular random $d$-hypergraphs for several values of $d$ obtained numerically and compare it with the theoretical expression, Eq.~\eqref{eq:Pk_proj}, using the approximation given in Eq.~\eqref{eq:Pdkappa}. We find excellent agreement between simulations and the theoretical prediction given by Eq.~\eqref{eq:Pk_proj}. We also compare the results with the theoretical curves in the sparse regime where the relation $k=d\kappa$ is valid. The plot illustrates how the overlap increases with the average hyperdegree $\mu_d$, and deviations from the sparse limit arise. 

 
\section{Overlap in random $d$-hypergraphs} \label{app:Overlap}

\subsection{Erd\H{o}s--R\'enyi $d$-hypergraphs} \label{app:Overlap_ER}
To quantify deviations from the sparse limit, we introduce the \emph{multiplicity} $R_{ij}$ as the number of $d$-hyperedges that simultaneously contain nodes $i$ and $j$. Note that this quantity is symmetric $R_{ij}=R_{ji}$. 

Given that there are $M^{(2)}_d=\binom{N-2}{d-1}$ possible hyperedges containing both $i,j$, and each is created independently with probability $p=\mu_d/\binom{N-1}{d}$, $R_{ij}$ follows a binomial distribution~\cite{Proj_ntwrk}
\begin{equation}\label{eq:p(oij)}
 P(R_{ij}=R)=\binom{M^{(2)}_d}{R}p^{R}(1-p)^{M^{(2)}_d-R}.
\end{equation}
A value $R_{ij}=0$ means that $i$ and $j$ are not connected in the projected network, while $R_{ij}=1$ corresponds to a single shared hyperedge. Values $R_{ij}\ge 2$ indicate \emph{overlap}, a situation where multiple hyperedges induce the same pairwise link in the projected network. The probability of overlap is therefore given by
\begin{equation} \label{Pov_ER}
 P_\text{overlap}\equiv P(R_{ij}\geq 2) = 1 - P(R_{ij}=0) -P(R_{ij}=1)=1 - (1-p)^{M^{(2)}_d} 
 - M^{(2)}_d \, p \, (1-p)^{M^{(2)}_d-1}.
\end{equation}

Let $O_i$ denote the event that node $i$ exhibits overlap. If we assume independence between overlap events for different neighbors $j\ne i$, the approximate probability that node $i$ exhibits overlap reads
\begin{equation}\label{eq:p_overlap_i}
 P(O_i) \;\approx\; 1-\bigl(1-P_\text{overlap}\bigr)^{N-1},
\end{equation}
which coincides with the expected fraction of nodes with overlap. Although weak correlations arise from hyperedges that simultaneously contain triples $(i,j,k)$, quadruples $(i,j,k,l)$, and so on, their effect vanishes as $N$ increases and we expect Eq.~\eqref{eq:p_overlap_i} to provide a good approximation in the large $N$ limit. 

\begin{figure}[t]
 \centering
 \includegraphics[width=0.45\linewidth]{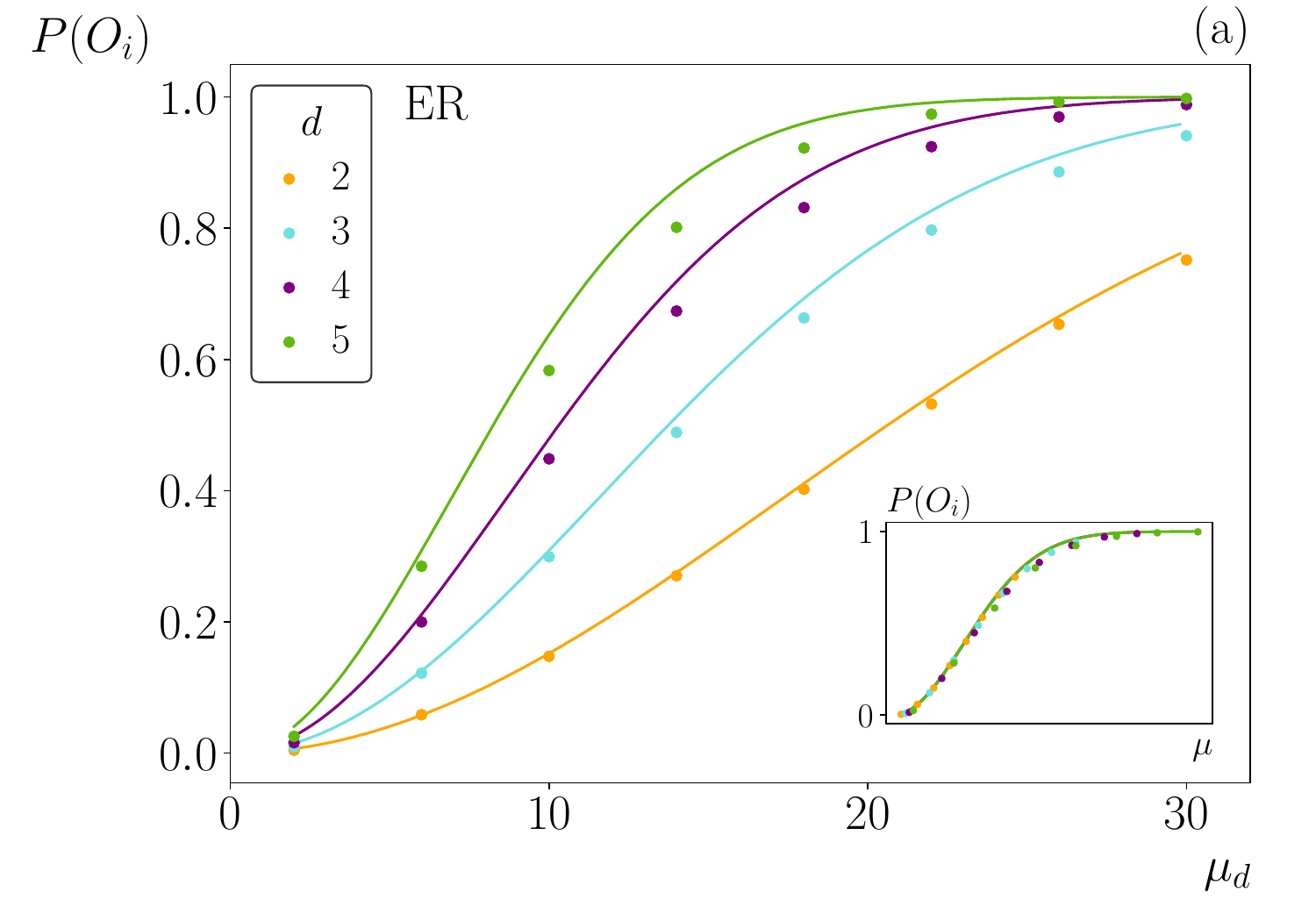}
 \includegraphics[width=0.45\linewidth]{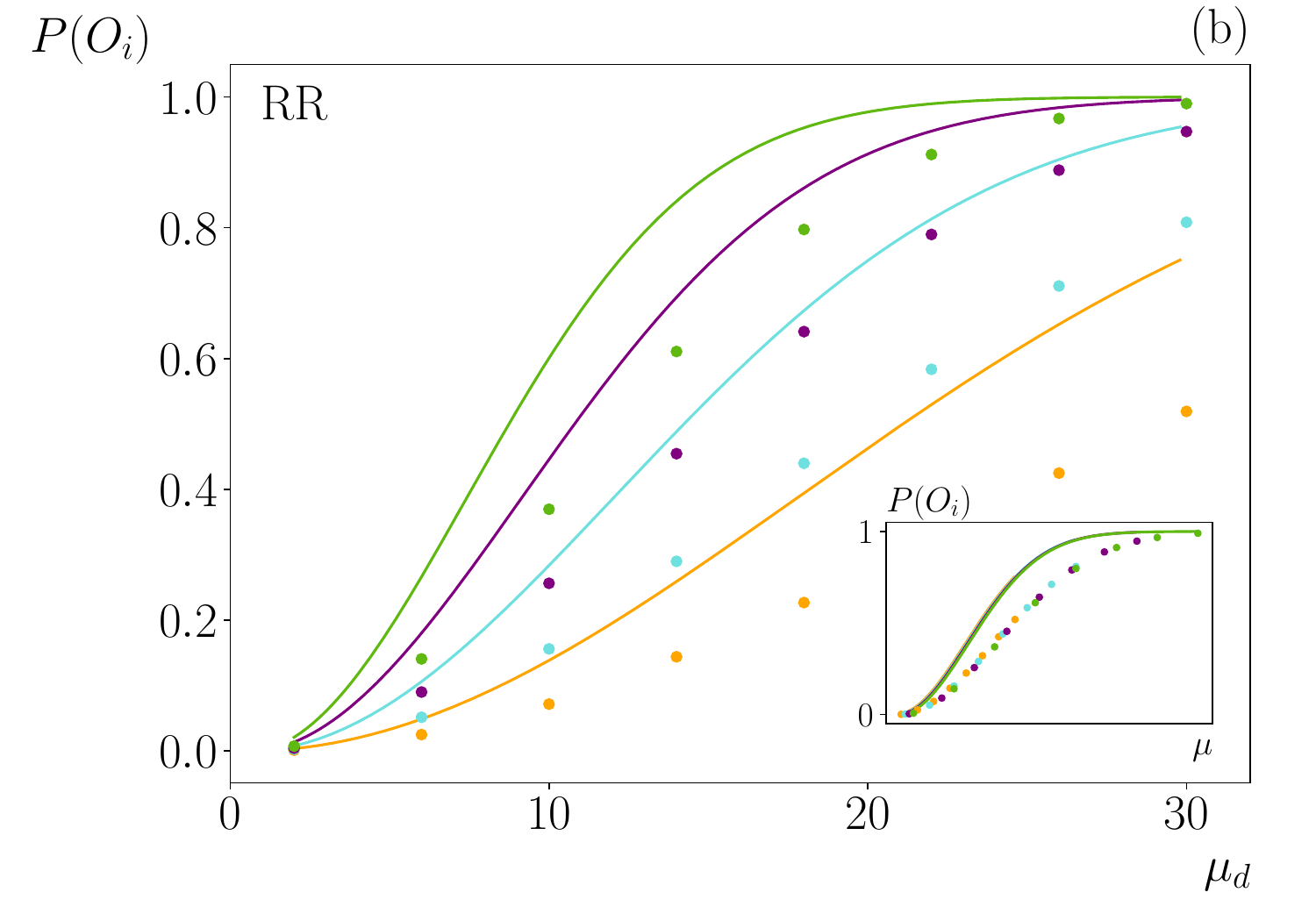}
 \caption{Fraction of nodes $P(O_i)$ with overlap in the projected network of (a) Erd\H{o}s--R\'enyi (ER) and (b) $\mu_d$-regular random (RR) $d$-hypergraphs versus the average $d$-hyperdegree $\mu_d$ for several values of $d$, as indicated in the legend. Symbols correspond to the results obtained from projected networks of $1000$ different hypergraphs generated with the algorithms detailed in Appendix~\ref{sec:app:HG_generation}. Lines represent the theoretical fraction of nodes $P(O_i)$ given in Eq.~\eqref{eq:p_overlap_i} together with Eqs.~(\ref{Pov_ER},\ref{Pov_RR}). Inset: same fraction $P(O_i)$ versus the average degree of the projected network, $\mu=d\,\mu_d$. All curves collapse. Legend is shown once for clarity. System size $N=1200$.}
 \label{fig:overlap}
\end{figure}

In Fig.~\ref{fig:overlap}, we plot Eq.~\eqref{eq:p_overlap_i} as a function of the average $d$-hyperdegree $\mu_d$ for several values of $d$. We find good agreement across all cases, with a trend of increasing deviations for larger $d$ due to correlations. As expected, given an average hyperdegree $\mu_d$, the fraction of nodes with overlap increases with $d$. Moreover, the inset shows the collapse of all curves when rescaled by the average degree of the projected network $\mu$.

Additionally, the overall degree of overlap can be assessed by considering all pairs of nodes. Since hyperedges are formed independently, the number $N_R$ of pairs $(i,j)$ that share exactly $R$ hyperedges follows a binomial distribution with single-event probability $P(R_{ij}=R)$ and expected value
\begin{equation}
 \langle N_R \rangle = \binom{N}{2}\, P(R_{ij}=R).
\end{equation}

This distribution allows us to compute the expected number of pairs of nodes that belong to more than one hyperedge
\begin{equation}
 \langle O^{(d)}\rangle = \sum_{R=2}^{M_d^{(2)}} \langle N_R^{(d)} \rangle,
\end{equation}
and the expected number of links in the projected network
\begin{equation}
 \langle E\rangle = E_d - \sum_{R=2}^{M_d^{(2)}} \langle N_R^{(d)}\rangle\,(R-1),
\end{equation}
where $E_d$ denotes the total number of pairwise links induced by all hyperedges before removing multiplicities. The correction term subtracts the $(R-1)$ redundant copies of each overlapped pair.

\subsection{$z$-regular random $d$-hypergraphs} \label{app:Overlap_RR}
In a $z$-regular random $d$-hypergraphs, every node participates in exactly $z$ hyperedges such that $z$ neighbor-slots of size $d$ are created. We can approximate the probability that a neighbor $j$ is selected for one slot as $p = d / (N-1)$, such that the probability that two nodes share $R{ij}=R$ hyperedges follows a binomial distribution,
\begin{equation} \label{eq:P_Rij_RR}
 P(R_{ij}=R)
 = \binom{z}{R}\, p^{\,R}\,\bigl(1-p\bigr)^{z-R}.
\end{equation}
The corresponding probability of overlap corresponding to the event $R_{ij}\ge 2$ is
\begin{equation} \label{Pov_RR}
 P_{\mathrm{overlap}}
 = P(R_{ij}\ge 2)= 1 - (1-p)^{z}
 - z \,p (1-p)^{z-1},
\end{equation}
allowing us to calculate the expected fraction of nodes with overlap again with Eq.~\eqref{eq:p_overlap_i}. 

In Fig.~\ref{fig:overlap}, we plot the fraction of nodes $P(O_i)$ with overlap as a function of the $d$-hyperdegree $z=\mu_d$ observing the same qualitatively phenomena exhibited by Erd\H{o}s--R\'enyi $d$-hypergraphs. However, deviatons from the theoretical curves Eqs.~(\ref{eq:p_overlap_i}, \ref{Pov_RR}) are more pronounced.

\section{Distribution of state configurations of hyperedges in complete $d$-hypergraphs}
\label{sec:app:hypergeometric-binomial}

 In this section we show that the hypergeometric distribution $\mathbf{H}(N_+,N_-,d,n)$ [Eq.~\eqref{eq:hyper_def}], which gives the probability of different state configurations of incident $d$-hyperedges in a complete $d$-hypergraph, can be approximated by the binomial distribution $\mathbf{B}(n,d;x)$ [Eq.~\eqref{eq:binomial}] when $N\gg d$. The key observation is that selecting a uniformly random incident $d$-hyperedge of a node is equivalent to drawing $d$ distinct nodes uniformly at random from the remaining $N-1$ nodes.

Let $N_\pm$ denote the numbers of nodes in state $\pm1$, with $N_+ + N_- = N$, and recall that the hyperdegree of each node is $M_d = \binom{N-1}{d}$. Considering a node $i$ with state $s_i=-1$, the number of $d$-hyperedges containing exactly $n$ positive and $d-n$ negative nodes is $\binom{N_+}{n}\binom{N_- -1}{d-n}$. This is calculated as the product of the number of different sets of $n$ distinct nodes that can be formed with $N_+$ positive nodes $\binom{N_+}{n}$, and the number of different sets of $d-n$ distinct nodes that are formed from the $N_{-}-1$ negative nodes $\binom{N_{-}-1}{d-n}$. Then, the hypergeometric distribution
\begin{equation}
\mathbf{H}(N_+,N_- -1,d,n)
= \frac{1}{M_d} \binom{N_+}{n}\binom{N_- -1}{d-n}
\label{eq:H-}
\end{equation}
gives the probability that a node $i$, with state $s_i=-1$, selects an incident $d$-hyperedge composed of $n$ nodes in state $+1$ and $d-n$ nodes in state $-1$.
Similarly, the probability that an incident $d$-hyperedge of a node $i$ with state $s_i=+1$ has $n$ and $d-n$ positive and negative nodes, respectively, is 
\begin{equation}
\mathbf{H}(N_+-1,N_-,d,n)
= \frac{1}{M_d} \binom{N_+ -1}{n}\binom{N_-}{d-n}.
\label{eq:H+}
\end{equation}
On the other hand, the probability of selecting $d$ random nodes without replacement and obtaining $n$ nodes in state $+1$ and $d-n$ nodes in state $-1$, given $N_\pm$, is
\begin{equation}
\tilde{\mathbf{H}}(N_+,N_--1,d,n)
= \binom{d}{n}\,
 \frac{(N_+)_{n}}{(N-1)_{n}}
 \frac{(N_--1)_{d-n}}{(N-n-1)_{d-n}},
\label{eq:H-nd}
\end{equation}
where we have defined the falling factorial $(N)_{n} \equiv N(N-1)\cdots(N-n+1)$. The terms $(N_+)_{n}/(N-1)_{n}$ and $(N_--1)_{d-n}/(N-n-1)_{d-n}$ give the probability of randomly selecting $n$ positive nodes and $d-n$ negative nodes, respectively, while the term $\binom{d}{n}$ accounts for the number of ways to choose these nodes. By applying the definition of the falling factorial, using the property $(N)_{n}= N!/[(N-n)!]$ and rearranging terms we obtain
\begin{equation}
\tilde{\mathbf{H}}(N_+,N_--1,d,n)
= \frac{\binom{N_+}{n}\binom{N_--1}{d-n}}{\binom{N-1}{d}},
\label{eq:P-nd_equiv}
\end{equation}
which precisely coincides with Eq.~\eqref{eq:H-}, showing that $\mathbf{H}(X,Y,d,n)=\tilde{\mathbf{H}}(X,Y,d,n)$. \\

We next show that, in the limit $N \gg d$, the hypergeometric distribution
$\mathbf{H}(N_+,N_-,d,n)$ can be approximated by the binomial distribution
$\mathbf{B}(n,d;x)$. 

Given that the falling factorial can be written as 
\begin{equation} (N)_n=\prod^{n-1}_{k=0}(N-k)=N^n\prod^{n-1}_{k=1}\left(1-\frac kN\right), \end{equation} 
one can Taylor expand the above product to first order in $n/N \le d/N \ll 1$ as
\begin{equation} 
\label{eq:app:taylor} 
 \prod^{n-1}_{k=1}\left(1-\frac kN\right)=1-\frac{n(n-1)}{2N}+O(N^{-2}). 
\end{equation}
Introducing the density of nodes in state $+1$, $x = N_+/N$ (so that $1-x = N_-/N$), and applying the approximation from Eq.~\eqref{eq:app:taylor} to every falling factorial of Eq.~\eqref{eq:H-nd} we obtain, after some algebra,
\begin{equation}
\mathbf{H}(N_+,N_- -1,d,n) = \tilde{\mathbf{H}}(N_+,N_- -1,d,n)
\simeq \binom{d}{n} x^{n}(1-x)^{d-n}
\left[
1 - \frac{
n\bigl[(2x^{2}+1)n - 2(1+x)d - 1\bigr] + x^{2} d(d+1)
}{2x(1-x)N}
\right],
\label{eq:H_binomial_correction}
\end{equation}
up to first order in $d/N$. In the $d/N \ll 1$ limit, the correction term vanishes and the hypergeometric distribution reduces to the binomial distribution
$\mathbf{B}(n,d;x)=\binom{d}{n}x^{n}(1-x)^{d-n}$. An analogous result holds for
$\mathbf{H}(N_+-1,N_-,d,n)$.

In summary, the hypergeometric distributions from Eqs.~(\ref{eq:H-}) and (\ref{eq:H+}) can be well approximated as
\begin{equation}
 \textbf{H}(N_+, N_{-}-1, d, n) \simeq\textbf{H}(N_{+}-1, N_{-}, d, n) \simeq \textbf{B}(n,d;x).
\end{equation}
The difference between the hypergeometric and the binomial distributions is that the first describes a process without replacement of nodes, making sure that all $d$ nodes in the group are different, whereas the second describes the same process but with replacement, where a node in the group can be selected more than once. Given that, in the $N \gg 1$ limit, the probability to select two or more times the same node is negligible (of order $1/N$), the two distributions become equivalent.

\section{Mean-field equation for the hypergraph-nonlinear voter model on the complete $d$-hypergraph}\label{app:Hdq}

The drift function gives the time evolution for $x$ in the thermodynamic limit. For the hypergraph-nonlinear voter model, the drift $A_d(x;q)\equiv T^+(x)-T^-(x)$ is obtained from Eqs.~\eqref{eq:T_global-1} and \eqref{eq:omega_CdH} as
\begin{equation} \label{eq:app:drift_def}
 A_d(x;q)
 =(1-x)\,\frac{\langle n^q \rangle}{d^q} 
 - x \,\frac{\langle (d - n)^q \rangle}{d^q},
\end{equation}
where $q>0$ is the nonlinearity parameter, $d$ is the order of the hyperedges and $\langle\cdot\rangle$ denotes the average over the binomial distribution $\textbf{B}(n;d,x)=\binom{d}{n}x^{n}(1-x)^{d-n}$, valid in the limit of a large system size $N\gg d$. 

For any real $q>0$, one can write the power of an integer $n$ as
\begin{equation}
 n^q = \sum_{k=0}^{\infty} S(q,k)\,\frac{n!}{(n-k)!},
 \label{eq:ndq_stirling_app}
\end{equation}
where $S(q,k)$ are the \textit{generalized Stirling numbers of the second kind}, still defined by Eq.~\eqref{eq:stirling2} for non-integer values of $q$. This formulation allows us to write down the averages of Eq.~\eqref{eq:app:drift_def} as
\begin{align}
 \left\langle n^q \right\rangle
 &= \sum_{k=0}^{d} S(q,k)\,\frac{d!}{(d-k)!}\,x^k,
 \label{eq:ndq_moment_app}
 \\
 \left\langle (d-n)^q \right\rangle
 &= \sum_{k=0}^{d} S(q,k)\,\frac{d!}{(d-k)!}\,(1-x)^k,
 \label{eq:dminusndq_moment_app}
\end{align}
where we have used
\begin{equation}
 \left\langle \frac{n!}{(n-k)!} \right\rangle
 = \sum_{n=0}^{d} \frac{n!}{(n-k)!}\,\mathbf{B}(n;d,x)
 = \frac{d!}{(d-k)!}\,x^k.
 \label{eq:falling_binomial_app}
\end{equation}

Substituting Eqs.~\eqref{eq:ndq_moment_app}–\eqref{eq:dminusndq_moment_app}
into Eq.~\eqref{eq:app:drift_def}, the drift becomes
\begin{align}
 A_d(x;q) = \frac{1}{d^{\,q}}
 \sum_{k=2}^{d} S(q,k)\,\frac{d!}{(d-k)!}
 \Big[ (1-x)\,x^{k} - x\,(1-x)^{k} \Big],
 \label{eq:Ad_before_fact_app}
\end{align}
where the sum starts from $k=2$ since the first two terms $k=0,1$ vanish. This drift is zero for $d=1$, whereas for $d>1$ can be rewritten as 
\begin{equation}
 A_d(x;q)=x(1-x)(2x-1)\,H_d(x;q),
 \label{eq:Adq}
\end{equation}
with the function $H_d(x;q)$ 
\begin{equation}\label{eq:Hqd_def_app}
 H_{d}(x;q)= 
 \begin{cases}
 \dfrac{1}{d^{\,q}}
 \displaystyle\sum_{k=2}^{d} 
 S(q,k)\, \frac{d!}{(d-k)!}(k-1)\,2^{-(k-2)}, 
 & \text{if } x= \dfrac{1}{2}, \\[1.1em]
 \dfrac{1}{d^{\,q}}
 \displaystyle\sum_{k=2}^{d} 
 S(q,k)\, \frac{d!}{(d-k)!}
 \dfrac{x^{\,k-1}-(1-x)^{\,k-1}}{2x-1}, 
 & \text{otherwise.}
 \end{cases}
\end{equation}

For arbitrary $q\neq1$ and $d>1$, the drift vanishes at least for three different values of $x$ that correspond to steady-state solutions of the system: the symmetric stationary state at $x=1/2$, and the two absorbing states at $x=0$ and $x=1$. To assess the possible existence of additional stationary solutions we introduce the auxiliary function
\begin{equation}
 \Delta_d(x;q)\equiv \frac{A_d(x;q)}{x(1-x)} = \frac{\mu_q(x)}{x}-\frac{\mu_q(1-x)}{1-x},
\end{equation}
where we have used $ \mu_q(x)\equiv \langle n^q \rangle/d^q$,
such that $\Delta_d(x;q)$ and $A_d(x;q)$ have the same sign in the interval $x\in[0,1]$ and $H_{d}(x;q)=\Delta_d(x;q)/(2x-1)$.

Since $n^q$ is an increasing function for any $q>0$, 
the average $\mu_q(x)$ grows monotonically with $x$. Hence $\mu_q(x)<\mu_q(1-x)$ for $x<1/2$ and the opposite for $x>1/2$. Moreover, the function $\mu_q(x)/x$ is monotonically increasing for $q>1$ and decreasing for $q<1$, which implies that $\Delta_d(x;q)$ has the same (resp. opposite) sign as $(2x-1)$ when $q>1$ (resp. $q<1$). Therefore, for all $x \in [0,1]$ we have that $H_{d}(x;q)>0$ when $q>1$, whereas $H_{d}(x;q)<0$ when $q<1$. Thus, there are no additional zeros of $A_d(x;q)$ in $x\in[0,1]$, and we conclude that $H_{d}(x;q)$ does not introduce any new real solutions to the dynamics.

The stability of the stationary states depends on $q$. For $q<1$, $H_d(x;q)<0$ and the symmetric state $x=1/2$ is linearly stable, while the absorbing states $x=0$ and $x=1$ are unstable. For $q>1$, one finds $H_d(x;q)>0$, implying that $x=0$ and $x=1$ are linearly stable absorbing states and that $x=1/2$ is unstable.


\section{Hypergraph-linear voter model: Weights distribution in the projected network of a $z$-regular random $d$-hypergraph}\label{sec:app:weights}

In a $z$-regular random $d$-hypergraph, all nodes have hyperdegree $\kappa_i = z$, $i=1,\dots,N$. Hence, according to Eq.~\eqref{eq:wij_VM}, the weight between $\oij$ simplifies to
\begin{equation}
 \oij =\frac{R_{ij}}{zd},
\end{equation}
where $R_{ij}$ is the number of $d$-hyperedges containing both nodes $i$ and $j$. Since $w_{ij}$ is a linear rescaling of $R_{ij}$, $w_{ij} \in \{ 0,\, \Delta w,\, 2\,\Delta w,\,\dots,\, z\Delta w \}$, with $\Delta w = 1/(z\,d)$, it follows the same distribution given in Eq.~\eqref{eq:P_Rij_RR}, $P\bigl( R_{ij}=R)=P\bigl( w_{ij}=R\,\Delta w)$.

We are interested in the distribution of weights among links that are actually present in the projected network, i.e. pairs with $R_{ij}\ge 1$. The conditioned probability on $R_{ij}\ge 1$ is given by
\begin{equation}
 P\bigl( w_{ij} = R\,\Delta w \,\big|\, R_{ij}\ge 1 \bigr)
 = P\bigl( R_{ij} = R|\, R_{ij}\ge 1 \bigr)=
 \frac{P(R_{ij}=R)}{1 - P(R_{ij}=0)}
 =
 \binom{d\,\kappa}{R}\frac{\displaystyle
 \, p^{\,R}\,
 \bigl(1-p\bigr)^{\,d\,\kappa-R}
 }{\displaystyle
 1 - \bigl(1-p\bigr)^{d\,\kappa}
 },
\end{equation}
with $p=1/(N-1)$ and $R = 1,2,\dots, d\,\kappa$, which allow us to estimate the probability that two nodes share more than one hyperedge, if they are connected, as
\begin{equation} \label{eq:Pov_ij}
 P(w_{ij}\ge 2\Delta w \, |\, R_{ij}\geq1)=P(R_{ij}\ge 2 \, |\, R_{ij}\geq1)= \dfrac{1-P(R_{ij}=1)-P(R_{ij}=0)}{1-P(R_{ij}=0)}.
\end{equation}
In our computer simulations, we have considered hyperedges of order up to $d=5$ and hyperdegrees up to $\kappa=15$, with system sizes of order $N\sim 10^3$--$10^4$. Equation~\eqref{eq:Pov_ij} for $d=5$ and $\kappa=15$ yields $P(w_{ij}\ge 2\Delta w \, |\, R_{ij}\geq1)\approx 3.66\times 10^{-2}$ for $N=10^3$ and $P(w_{ij}\ge 2\Delta w\, |\,R_{ij}\geq1)\approx 3.70\times 10^{-3}$ for $N=10^4$. Even in the most demanding combination of parameters the probability that a given pair $(i,j)$ participates
in more than one hyperedge is extremely small. Therefore, the weight distribution reduces to the degree-homogeneous case for the range of values of the parameters used in our computer simulations, namely
\begin{equation} \label{eq:wij_VM_RR}
 \oij \approx \begin{cases}
 \dfrac{1}{z d}& \text{if $i$ and $j$ are connected}, \\
0& \text{otherwise}.
 \end{cases}
\end{equation} 
In other words, for the hypergraph-linear voter model on a $z$-random regular $d$-hypergraph, the equivalent PW dynamics on the weighted projected network coincides in the sparse limit with the VM on the projected network. The projected network is not exactly a standard $z d$-random regular network. Instead, it presents structural differences arising from the clustering induced by the hypergraph. This clustering induces correlations that decrease the value of the plateau~\cite{Gleeson_corr}, as discussed in Sec.~\ref{sec:Map_VM}.

\section{Pair approximation for a pairwise dynamics on complex networks with static weights}
\label{app:PA}

In this section we develop a pair approximation (PA) for the pairwise dynamics associated to the hypergraph-linear voter model. Due to the linearity of the social impact function for this particular case [$f(\phi)=\phi$], the weights in the projected network turn to be constant over time (see Sec.~\ref{sec:Map_VM}), allowing analytical treatment. Within the PA, we shall obtain two coupled differential equations for the time evolution of the fraction of nodes $x$ in state $+1$, and the density of active links $\rho$. These two macroscopic quantities, $x$ and $\rho$, characterize very well the ordering properties of the system in voter models \cite{Ramirez2024}. 

In general, given the rules of the PW dynamics, the possible weights $\omega_{ij}$ can take a discrete set of values that we denote by $\{\omega_\alpha\}_{\alpha=1,2,\dots}$. Let us denote by $P(k_1,k_2,\dots)\equiv P(\vec k)$ the fraction of nodes that are connected to $k_\alpha$ links of weight $\omega_\alpha$, ($\alpha=1,2,\dots$). By definition, we have the normalization condition 
\begin{equation}\label{eq:normaPk}
\sum_{k_1,k_2,\dots}P(k_1,k_2,\dots)\equiv \sum_{\vec k}P(\vec k)=1.
\end{equation}
The usual node degree distribution is obtained from
\begin{equation}
P(k)=\sum_{\vec k}\delta\left(\sum_\alpha k_\alpha-k\right)P(\vec k).
\end{equation}

If a node $i$ is connected to a total of $k$ links, out of which $k_\alpha$ are links of weight $\omega_\alpha$, we have the following identities:
\begin{subequations}\label{eq:normalization_ap}
 \begin{align}
 k&=\sum_{\alpha} k_{\alpha}, \label{eq:normalizations}\\
 1&=\sum_{\alpha} \omega_{\alpha} k_{\alpha},
 \label{eq:k-topo}
 \end{align} 
\end{subequations}
the second equality arising from the normalization condition $\sum_{j}\omega_{ij}=1$, after splitting the $k$ links of node $i$ into its different weights $k_1,k_2,\dots$.

In the following, we derive evolution equations for the evolution of $x$ and $\rho$ on the weighted projected network. For the sake of simplicity in the calculations, we introduce the variables $\sigma_+=x$ and $\sigma_-=1-x$. We follow closely the approach developed in Ref.~\cite{Vazquez_PA} for the VM on uncorrelated complex networks, considering a general case scenario in which the interacting neighbors are not chosen uniformly at random, as in the standard VM \cite{Vazquez_PA}, but with a probability given by the weights of the projected network. 

In a time step $\Delta t=1/N$ of the pairwise dynamics, a node $i$ chosen at random copies the state of one of its neighbors $j$ with probability $\omega_{ij}$. The probability that node $i$ has state $s_i=s$ is $\sigma_s$, and the probability that it is connected to $k_\alpha$ neighbors with weights $\omega_\alpha$, ($\alpha=1,2,\dots$) is $P(\vec k)$.

We consider that a neighbor $j$ of the focal node $i$ is in the opposite state $s_j=-s$ (link $ij$ is active) with a probability $P_{-s|s}$ that only depends on the state $s$ of $i$, and not on the state of other neighbors of $i$ (pair approximation). We are also assuming that $P_{-s|s}$ neither depends on the weight of the link between $i$ and $j$, nor on the degrees of $i$ and $j$, that is, we consider a PA that is homogeneous in weights and degrees. Then, the probability that $n_{\alpha}$ of the $k_{\alpha}$ links of weight $\omega_{\alpha}$ ($\alpha=1,2,\dots$) are active is given by the product of the binomial distributions $\prod_{\alpha} \textbf{B}(n_{\alpha},k_{\alpha},P_{-s|s})$ with single event probability $P_{-s|s}$ that a link is active. The focal node $i$ flips when the chosen neighbor is connected to an active link, which happens with probability $\sum_{\alpha} \omega_{\alpha} n_{\alpha}$, after which $\sigma_s$ changes by $\Delta \sigma_s = -s/N$. Thus, assembling all these factors and summing over all possible configurations of links $\Vec{k}$ and active links $\Vec{n}$, the time evolution of $\sigma_{s}$ can be written as
\begin{eqnarray}
 \frac{d\sigma_{s}(t)}{dt} &=& \sum_{s=\pm} \frac{\sigma_{s}}{1/N} \sum_{\vec k} P(\vec k) \sum_{\vec n} \prod_{\alpha} \textbf{B}(n_{\alpha},k_{\alpha},P_{-s|s}) \sum_{\alpha'} \omega_{\alpha'} n_{\alpha'} \,\Delta\sigma_{s} \nonumber \\ 
 &=& -\sum_{s=\pm} s \, \sigma_{s} \sum_{\vec k} P(\vec k) \sum_{\alpha'} \omega_{\alpha'} \langle n_{\alpha'} \rangle, 
 \label{eq:dsdt-PA}
\end{eqnarray}
where $\sum_{\vec n}\equiv\sum_{n_1,n_2,\dots}$, and we have introduced $\langle n_{\alpha} \rangle$ as the first moment of the binomial distribution $\textbf{B}(n_{\alpha},k_{\alpha},P_{-s|s})$, namely $\langle n_{\alpha} \rangle=k_\alpha P_{-s|s}$. Note that we are using the fact that the weights $\omega_{\alpha}$ are constant, independent of $n_{\alpha}$, which only happens when the social impact function is linear. The conditional probability $P_{-s|s}$ can be estimated as the ratio between the number of links from nodes in state $s$ to nodes in state $-s$ ($N_{s \to -s} = \mu N \rho/2$) and the total number of links coming out from nodes in state $s$ ($N_{s \to \pm} = \mu N \sigma_s$), i.e., $P_{-s|s} = \rho/(2 \sigma_s)$, where $\mu = \sum_{k} k P(k)$ is the mean degree of the network. Thus, the first two moments of the binomial distribution $\textbf{B}(n_{\alpha},k_{\alpha},P_{-s|s})$ are 
\begin{eqnarray}
 \langle n_{\alpha} \rangle = \frac{k_{\alpha} \rho}{2 \sigma_s},\quad\quad
 \langle n_{\alpha}^2 \rangle = \frac{k_{\alpha} \rho}{2 \sigma_s} + \frac{k_{\alpha}(k_{\alpha}-1) \rho^2}{4 \sigma_s^2}.
 \label{eq:moments}
\end{eqnarray}
Replacing the expression for $\langle n_{\alpha} \rangle$ into Eq.~(\ref{eq:dsdt-PA}), and using the normalization conditions, Eqs.~(\ref{eq:normaPk},\ref{eq:normalization_ap}), we obtain $d\sigma_{s}(t)/dt=0$, and thus
\begin{eqnarray}
 \frac{dx}{dt}=0.
 \label{eq:dsdt-PA-1}
\end{eqnarray}
Therefore, the fraction of nodes in each state is conserved under the PW dynamics on the projected network.

We now derive a rate equation for the evolution of $\rho$ following the same steps as for $\sigma_s$ above, where now the change in $\rho$ in a flipping event is $\Delta \rho=2(k-2n)/(\mu N)=\sum_{\alpha} 2(k_{\alpha}-2n_{\alpha})/(\mu N)$. Here we have considered that the number of active links connected to the focal node changes from $n$ ($s \to -s$ links) to $k-n$ ($-s \to s$ links) when the node flips. Then, the average change in $\rho$ in a time step can be written as
\begin{eqnarray}
 \label{eq:drhodt-PA} 
 \frac{d\rho(t)}{dt} &=& \sum_{s=\pm} \frac{\sigma_{s}}{1/N} \sum_{\Vec{k}} P(\Vec{k}) \sum_{\Vec{n}} \prod_{\alpha} \textbf{B}(n_{\alpha},k_{\alpha},P_{-s|s}) \sum_{\alpha'} \omega_{\alpha'} n_{\alpha'} \sum_{\alpha''} \frac{2(k_{\alpha''} - 2n_{\alpha''})}{\mu N} \nonumber \\
 &=& \frac{2}{\mu} \sum_{s=\pm} \sigma_{s} \sum_{\Vec{k}} P(\Vec{k}) 
 \sum_{\alpha'}\sum_{\alpha''} \omega_{\alpha'}\left\langle n_{\alpha'} (k_{\alpha''}-2n_{\alpha''}) \right\rangle \nonumber \\
 &=& \frac{2}{\mu} \sum_{s=\pm} \sigma_{s} \sum_{\Vec{k}} P(\Vec{k}) 
 \Bigg\{ 
 \sum_{\alpha'} \sum_{\alpha''} \omega_{\alpha'} \langle n_{\alpha'} \rangle k_{\alpha''} - 2 \sum_{\alpha'} \omega_{\alpha'} \langle n_{\alpha'}^2 \rangle - 2 \sum_{\alpha'} \sum_{\alpha'' \ne \alpha'} \omega_{\alpha'} \langle n_{\alpha'} \rangle \langle n_{\alpha''} \rangle \Bigg\}. 
\end{eqnarray}
Replacing into Eq.~(\ref{eq:drhodt-PA}) the first and second moments from Eq.~(\ref{eq:moments}) we obtain
\begin{eqnarray}
 \frac{d\rho(t)}{dt} = \frac{\rho}{\mu} \sum_{s=\pm} \sum_{\Vec{k}} P(\Vec{k}) 
 \Bigg\{ \sum_{\alpha'} \sum_{\alpha''} \omega_{\alpha'} k_{\alpha'} k_{\alpha''} - 2 \sum_{\alpha'} \omega_{\alpha'} \left[ k_{\alpha'} + \frac{k_{\alpha'}(k_{\alpha'}-1) \rho}{2 \sigma_{s}} \right] - 2 \sum_{\alpha'} \sum_{\alpha'' \ne \alpha'} \frac{\omega_{\alpha'} k_{\alpha'} k_{\alpha''} \rho}{2 \sigma_{s}} \Bigg\}. 
 \label{eq:drhodt-PA-1}
\end{eqnarray}
Using the conditions Eqs.~(\ref{eq:normalization_ap}), the first term inside the curly brackets becomes $k$, while the second and third terms combined give $-2[1+(k-1)\rho]/(2\sigma_s)$. Plugging these expressions into Eq.~(\ref{eq:drhodt-PA-1}) and performing the sums over $s$ and $\vec k$ we finally arrive at
\begin{eqnarray}
 \frac{d\rho}{dt} = \frac{2\rho}{\mu} \left[ (\mu-1) \left(1-\frac{\rho}{2 x (1-x)} \right) - 1 \right]. 
 \label{eq:drhodt-PA-2}
\end{eqnarray}
Equations~(\ref{eq:dsdt-PA-1}) and (\ref{eq:drhodt-PA-2}) formed a closed system of coupled equations that describe the approximate evolution of the macroscopic variables $x$ and $\rho$, quoted in Eqs.~\eqref{eq:rate_eq_PA} of the main text. These equations turn to be the same as those of the PA for the VM on uncorrelated networks \cite{Vazquez_PA}, corresponding to uniform weights $\omega_{ij}=1/k_i$ for every neighbor $j$ of $i$ (see Sec.~\ref{sec:general_pw}). This result highlights the fact that, within the PA, the specific distribution of weights seem to be irrelevant for an imitation dynamics on networks.

\section{Fixation times for the hypergraph-nonlinear voter model in Erd\H{o}s--R\'enyi $d$-hypergraphs}\label{app:NLVM_HG}

\begin{figure*}[t]
 \centering
 \includegraphics[width=0.32\textwidth]{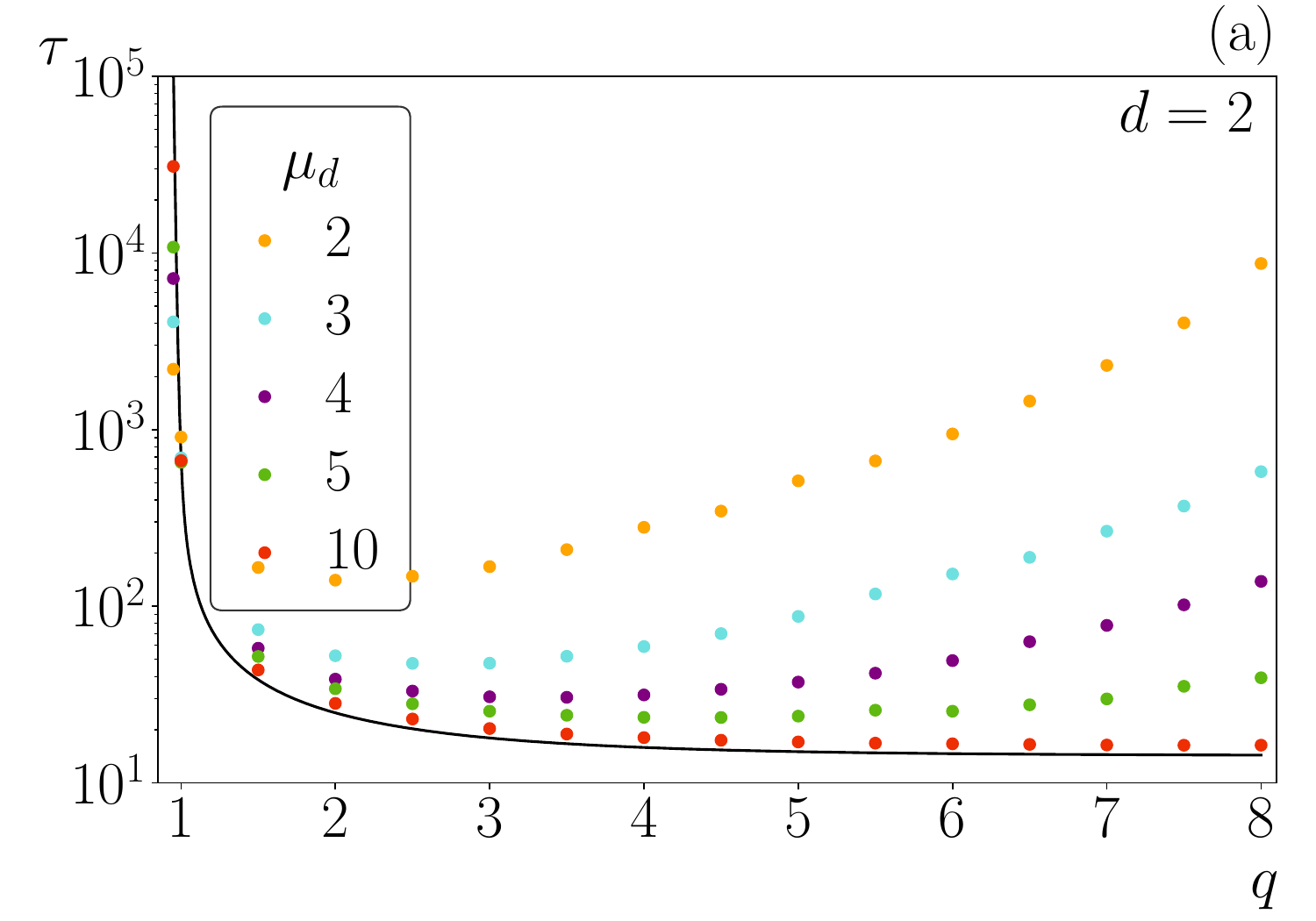}
 \includegraphics[width=0.32\textwidth]{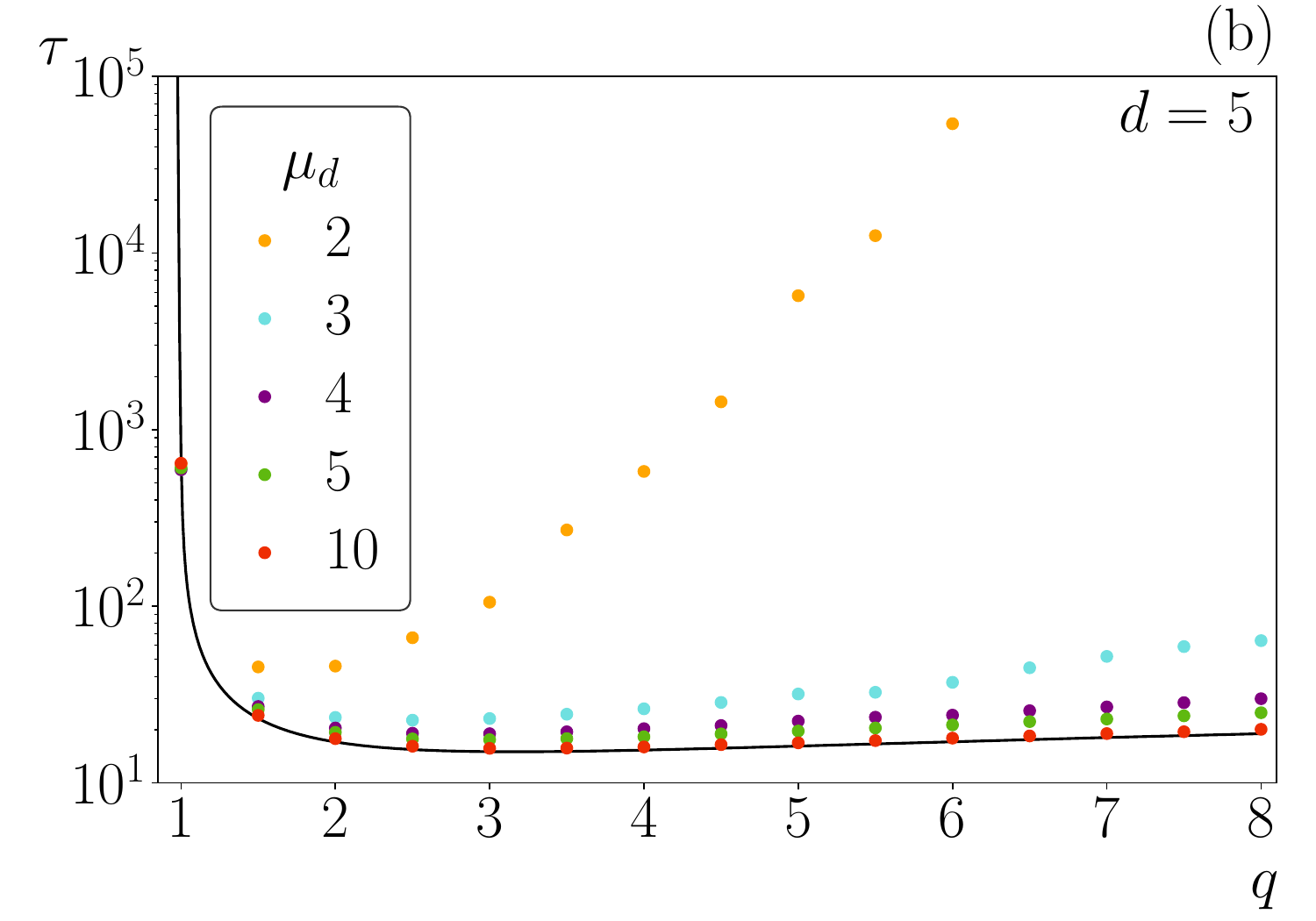}
 \includegraphics[width=0.32\textwidth]{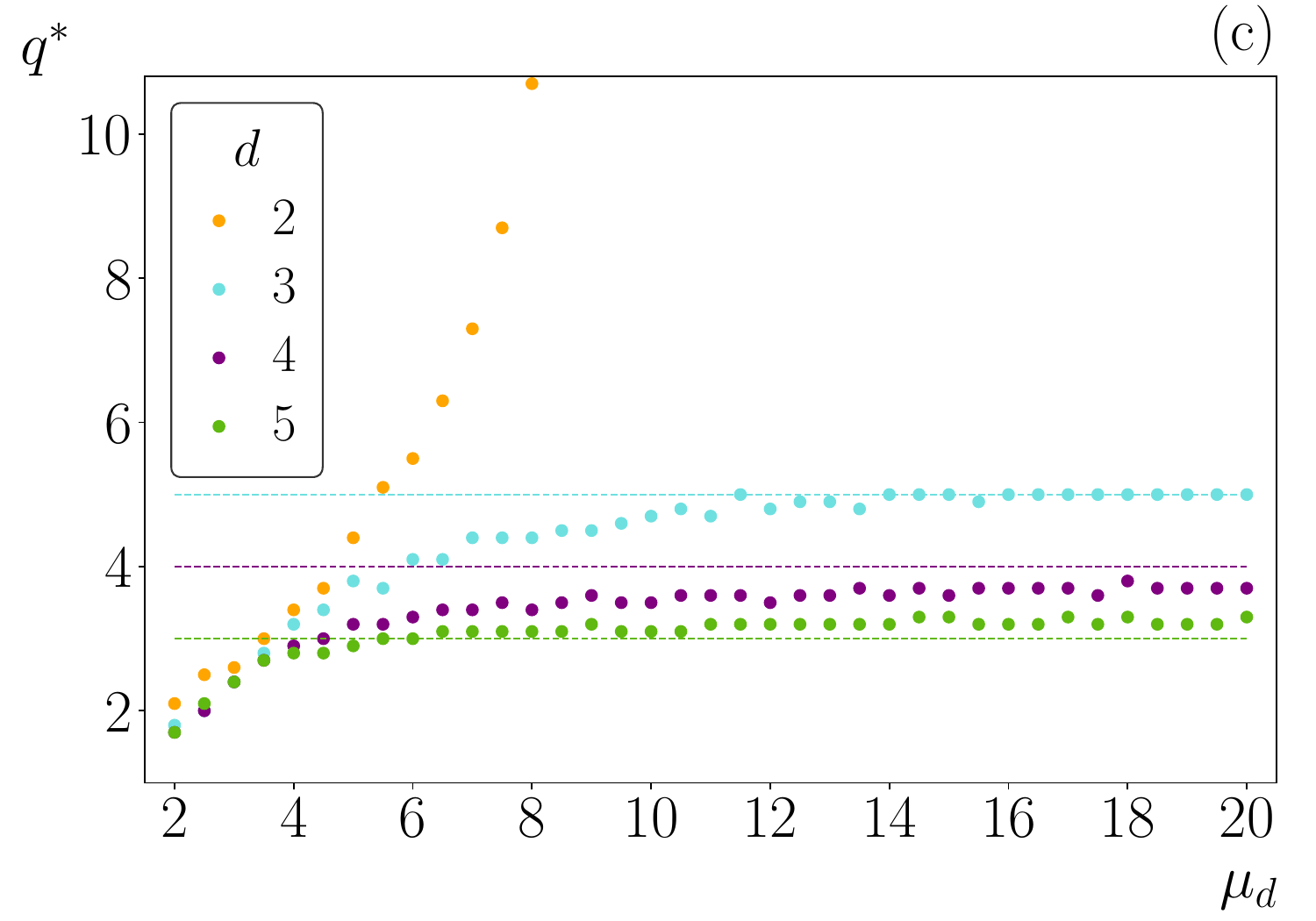}
 \caption{Fixation time $\tau$ versus the nonlinearity $q$ for Erd\H{o}s--R\'enyi $d$-hypergraphs for $d=2$ (a) and $d=5$ (b) for several values of the average hypedegree $\mu_d$. Symbols correspond to computer simulations for $N=1000$ while the black lines correspond to the theoretical solution on complete $d$-hypergraphs derived in Ref.~\cite{Min2025}. (c) Optimal nonlinearity $q^*$ versus the average hyperdegree $\mu_d$ for several values of $d$, as indicated in legend.}
 \label{fig:tau_HG}
\end{figure*}

We study the effects of hypergraph topology on the fixation time $\tau$ for arbitrary values of $q$, comparing our results with those reported for the group-driven voter model of Ref.~\cite{Min2025}, which were restricted to integer values of $q$ and complete $d$-hypergraphs. Figure~\ref{fig:tau_HG} shows $\tau$ as a function of the nonlinear parameter $q$ for Erd\H{o}s--R\'enyi $d$-hypergraphs and several values of the average $d$-hyperdegree $\mu_d$. We observe the existence of an optimal value $q^*$ that minimizes the fixation time, consistent with the findings of Ref.~\cite{Min2025} for the complete $d$-hypergraphs. This value $q^*$ depends on the average $d$-hyperdegree $\mu_d$. We find two qualitatively different behaviors depending on the group size $d$.

For $d=2$, we observe a qualitative change with $\mu_d$: $\tau$ is non-monotonic with a well-defined minimum value $q^*$ for small $\mu_d$, whereas the minimum shifts to $q\to\infty$ and $\tau$ decreases monotonically for large $\mu_d$. This indicates that topological heterogeneity strongly influences the ordering dynamics when interactions involve small groups. 

For $d=5$ the impact of heterogeneity is much weaker. The curves of $\tau$ for different $\mu_d$ display a qualitatively similar shape, and the main effect of increasing $\mu_d$ is simply to reduce the overall fixation time. This suggests that, for larger group sizes, the dynamics approaches the mean-field limit and the influence of heterogeneity diminishes. In both cases, the results converge to those of Ref.~\cite{Min2025} in the limit $\mu_d \to \infty$, confirming that the complete $d$-hypergraph is recovered asymptotically. We plot the optimal nonlinearity $q^*$ as a function of $\mu_d$, illustrating these two different regimes, in Fig.~\ref{fig:tau_HG} (c).

\end{document}